\documentclass[journal]{IEEEtran}

\usepackage{amsfonts}
\usepackage{amssymb}
\usepackage{amsthm}
\usepackage{amsmath,amsfonts,amssymb}
\usepackage[dvips]{graphicx}
\usepackage{subfigure}
\usepackage{verbatim}
\usepackage{setspace}
\usepackage{bm}
\usepackage{algorithmic} 
\usepackage[ruled,vlined]{algorithm2e}
\usepackage{cite}

\usepackage{changepage}
\usepackage{pdfpages}
\usepackage{color}
\newcommand*{\dif}{\mathop{}\!\mathrm{d}}

\allowdisplaybreaks
\IEEEoverridecommandlockouts

\begin{document}
\title{Modeling and Performance Analysis for Movable Antenna Enabled Wireless Communications}
\author{\normalsize Lipeng Zhu, ~\IEEEmembership{\normalsize Member,~IEEE,}
		Wenyan Ma,~\IEEEmembership{\normalsize Student Member,~IEEE,}
		and Rui Zhang,~\IEEEmembership{\normalsize Fellow,~IEEE}
	\thanks{This work is supported in part by Ministry of Education, Singapore under Award T2EP50120-0024, Advanced Research and Technology Innovation Centre (ARTIC) of National University of Singapore under Research Grant R-261-518-005-720, and The Guangdong Provincial Key Laboratory of Big Data Computing. (\textit{Corresponding author: Lipeng Zhu and Rui Zhang})}
	\thanks{L. Zhu and W. Ma are with the Department of Electrical and Computer Engineering, National University of Singapore, Singapore 117583 (e-mail: zhulp@nus.edu.sg, wenyan@u.nus.edu).}
	\thanks{R. Zhang is with School of Science and Engineering, Shenzhen Research Institute of Big Data, The Chinese University of Hong Kong, Shenzhen, Guangdong 518172, China (e-mail: rzhang@cuhk.edu.cn). He is also with the Department of Electrical and Computer Engineering, National University of Singapore, Singapore 117583 (e-mail: elezhang@nus.edu.sg). }
}

\maketitle



\begin{abstract}
	In this paper, we propose a novel antenna architecture called movable antenna (MA) to improve the performance of wireless communication systems. Different from conventional fixed-position antennas (FPAs) that undergo random wireless channel variation, the MAs with the capability of flexible movement can be deployed at positions with more favorable channel conditions to achieve higher spatial diversity gains. To characterize the general multi-path channel in a given region or field where the MAs are deployed, a field-response model is developed by leveraging the amplitude, phase, and angle of arrival/angle of departure (AoA/AoD) information on each of the multiple channel paths under the far-field condition. Based on this model, we then analyze the maximum channel gain achieved by a single receive MA as compared to its FPA counterpart in both deterministic and stochastic channels. First, in the deterministic channel case, we show the periodic behavior of the multi-path channel gain in a given spatial field, which can be exploited for analyzing the maximum channel gain of the MA. Next, in the case of stochastic channels, the expected value of an upper bound on the maximum channel gain of the MA in an infinitely large receive region is derived for different numbers of channel paths. The approximate cumulative distribution function (CDF) for the maximum channel gain is also obtained in closed form, which is useful to evaluate the outage probability of the MA system. Moreover, our results reveal that higher performance gains by the MA over the FPA can be acquired when the number of channel paths increases due to more pronounced small-scale fading effects in the spatial domain. Numerical examples are presented which validate our analytical results and demonstrate that the MA system can reap considerable performance gains over the conventional FPA systems with/without antenna selection (AS), and even achieve comparable performance to the single-input multiple-output (SIMO) beamforming system. 
\end{abstract}
\begin{IEEEkeywords}
	Movable antenna (MA), spatial diversity, field response, performance analysis.
\end{IEEEkeywords}

%
\IEEEpeerreviewmaketitle

\section{Introduction}
\IEEEPARstart{I}{n} the process of wireless systems evolution, larger capacity and higher reliability have always been the main objectives to pursue. Multiple-input multiple-output (MIMO) or multi-user/multi-antenna communication has been a key enabling technology in this endeavor, which lifts the veil on the new degrees of freedom (DoFs) in the spatial domain for improving the communication performance \cite{telatar1999capacity,Paulraj2004Anover,Stuber2004broadb}. With the current trend and future expectation of wireless communication systems migrating to higher frequency bands, such as millimeter-wave (mmWave) and terahertz (THz) bands, the decreasing wavelength results in smaller antenna size, which renders the MIMO system to be of larger scale (a.k.a. massive MIMO) in order to compensate for the more severe propagation loss \cite{Larsson2014massive,Ning2023THzbeam,Shao2022target,Wan2021HoloRIS}. Compared to conventional MIMO, massive MIMO is able to exploit the spatial correlation of large antenna arrays for attaining higher array gains and mitigating the multi-user interference more effectively \cite{Wan2021HoloRIS,zeng2016millim,zhu2019millim,Ning2020beamform}.

However, since the antennas are deployed at fixed positions, conventional MIMO and massive MIMO cannot fully explore the spatial variation of wireless channels in a given receive area or receive field. To overcome this limitation, we propose a new antenna architecture, namely movable antenna (MA), for exploiting the DoFs in the continuous spatial domain more efficiently. Different from conventional fixed-position antennas (FPAs), the positions of MAs can be flexibly adjusted in a spatial region for improving the channel condition and enhancing the communication performance. An alternative way to exploit spatial DoFs is the antenna selection (AS) technique for FPA systems. However, to achieve higher diversity orders for wireless transmission, the AS system requires increasingly more antenna elements, thus resulting in increased cost \cite{Molisch2004MIMOsys}. In contrast, the MA system can reap the full spatial diversity with much fewer or even one single antenna moving in a given region. Furthermore, an MA possesses the flexibility to move continuously in a three-dimensional (3D) space, enabling it to fully exploit the spatial channel variations in all possible directions. However, FPA systems, with or without using AS, face the limitation of discrete antenna positions within a one-dimensional (1D) line or two-dimensional (2D) surface only. The above advantages of the MA system allow for utilizing wireless channel spatial DoFs more efficiently, thus leading to improved communication performance \cite{zhu2023MAMag,ma2022MAmimo,zhu2023MAmultiuser}.

It is worth noting that the aforementioned performance improvement of MA systems relies on the capability of mechanically moving the antennas to achieve more favorable channel conditions. For wireless communication systems where the channels experience fast variation over time, the performance gains of MAs are difficult to be acquired due to the limited antenna-moving speed. Nonetheless, there are many communication scenarios where the wireless channels vary slowly over time due to the limited mobility of wireless terminals. For example, with the development of Internet of things (IoT), future wireless networks need to support various machine-type communication (MTC) devices in e.g., smart cities, automated industries, smart homes, etc. as well as a variety of wireless sensors \cite{shaik2021comprehensive}. These IoT devices are usually deployed in confined areas with fixed positions and have low or even no mobility, and their surrounding environments may change slowly \cite{Beyene2017narrowIoT,Dowhuszko2016delayMTC,Hsieh2018LTEmachine,Hyder2014SmartGrid}. In such scenarios, MA becomes a viable solution for exploiting the wireless channel spatial diversity to improve the communication performance, especially when traditional time and frequency diversities are unavailable, e.g., in narrow-band IoT (NB-IoT), long-range (LoRa), and machine-to-machine (M2M) communications.

The implementation of MA system resembles that of the existing distributed antenna system (DAS) \cite{Castanheira2010distri,Heath2013Acurrent,Cui2019greend}, while only local movement of MAs (in the order of several wavelengths) is required by connecting each MA to the radio-frequency (RF) chain via a flexible cable. MA systems have been implemented in practice for different applications. For example, the authors in \cite{Zhuravlev2015experi} showed a prototype of MA-enabled multi-static radar, where the transmit and receive antennas are connected to the vector network analyzer and can be moved with the aid of linear drivers. In \cite{Basbug2017design}, a reconfigurable antenna array was designed, where the array elements can move along a semicircular path with the aid of stepper motors. The authors in \cite{Do2021reconf} and \cite{Do2022TeMIMO} proposed a reconfigurable uniform linear array (ULA) architecture which enables the mechanical rotation of the antenna array for attaining the maximum capacity of line-of-sight (LoS) MIMO transmission. Considering the integration of inertial navigation system (INS) and global navigation satellite system (GNSS), the authors in \cite{Li2022movingAnte} proposed to employ a moving antenna for providing reliable and consistent state estimation under low-dynamic scenarios. Besides, for a pseudolite indoor localization system, a movable Rx antenna was utilized to increase the positioning accuracy by leveraging the Doppler shift \cite{Sakamoto2011movable}. Moreover, the authors in \cite{eliyahu2022single} proposed to use a single MA for direction finding (DF). However, the above works mainly focus on the application of MA in radar, navigation, localization, or LoS transmission systems \cite{Zhuravlev2015experi, Do2021reconf,Do2022TeMIMO, Basbug2017design, Li2022movingAnte,Sakamoto2011movable,eliyahu2022single}.

The employment of MAs in wireless communication systems has been recently investigated from various perspectives, such as spatial diversity, spatial multiplexing, and interference mitigation \cite{zhu2023MAMag,ma2022MAmimo,zhu2023MAmultiuser,Wong2021fluid,Wong2022fluid}. In \cite{zhu2023MAMag}, an overview on the application scenarios, technical potentials, practical challenges, and promising solutions for MA-aided wireless communications was presented. In \cite{ma2022MAmimo}, an MA-aided MIMO communication system was considered, where the positions of multiple MAs at the transmitter (Tx) and receiver (Rx) were jointly designed for maximizing the channel capacity. Besides, the MA-enhanced multiuser communication was investigated in \cite{zhu2023MAmultiuser} by jointly optimizing the positions of MAs and the transmit power of users, as well as the receive combining matrix at the base station (BS) in the uplink. The results in \cite{ma2022MAmimo} and \cite{zhu2023MAmultiuser} showed that the antenna movement within a sub-wavelength distance can significantly change the channel matrices/vectors due to the pronounced small-scale fading in spatial regions. In addition, the authors in \cite{Wong2021fluid} proposed a novel fluid antenna system (FAS) with controllable antenna position, which can be regarded as a particular way of implementing MAs. In addition, a fluid antenna multiple access (FAMA) system was proposed in \cite{Wong2022fluid} to support transceivers with different channel conditions simultaneously. However, the discrete port-based channel model employed in \cite{Wong2021fluid} and \cite{Wong2022fluid} cannot completely characterize and fully exploit the wireless channel spatial variation in continuous regions.

Given the above technical advantages and potential applications of MA, we aim to investigate in this paper the channel modeling and performance analysis for MA-enabled communication systems. To characterize the general multi-path channel in a given region or field where the MAs are deployed, a field-response model is developed by leveraging the amplitude, phase, and angle of arrival/angle of departure (AoA/AoD) information on each of the multiple channel paths under the far-field condition. Based on the proposed channel model, we then analyze the maximum channel gain achieved by a single receive MA as compared to its FPA counterpart in both deterministic and stochastic channels. In the deterministic channel case, we show the periodic behavior of the multi-path channel gain in a given spatial field. In the case of stochastic channels, the expected value of an upper bound on the maximum channel gain of the MA in an infinitely large receive region is derived for different numbers of channel paths. The approximate cumulative distribution function (CDF) for the maximum channel gain is also obtained in closed form. Moreover, our results reveal that higher performance gains by the MA over the FPA can be acquired when the number of channel paths increases due to more pronounced small-scale fading effects in the spatial domain. Extensive simulations are carried out to verify our analytical results and compare the performance of the proposed MA-enabled communication system with conventional FPA systems based on both the analytical channel model and the practical channel model suggested by the Third Generation Partnership Project (3GPP). The results demonstrate that the MA system can reap considerable performance gains over the FPA systems with/without AS, and even achieve comparable performance to the single-input multiple-output (SIMO) beamforming system.

\begin{figure*}[t]
	\begin{center}
		\includegraphics[width=14 cm]{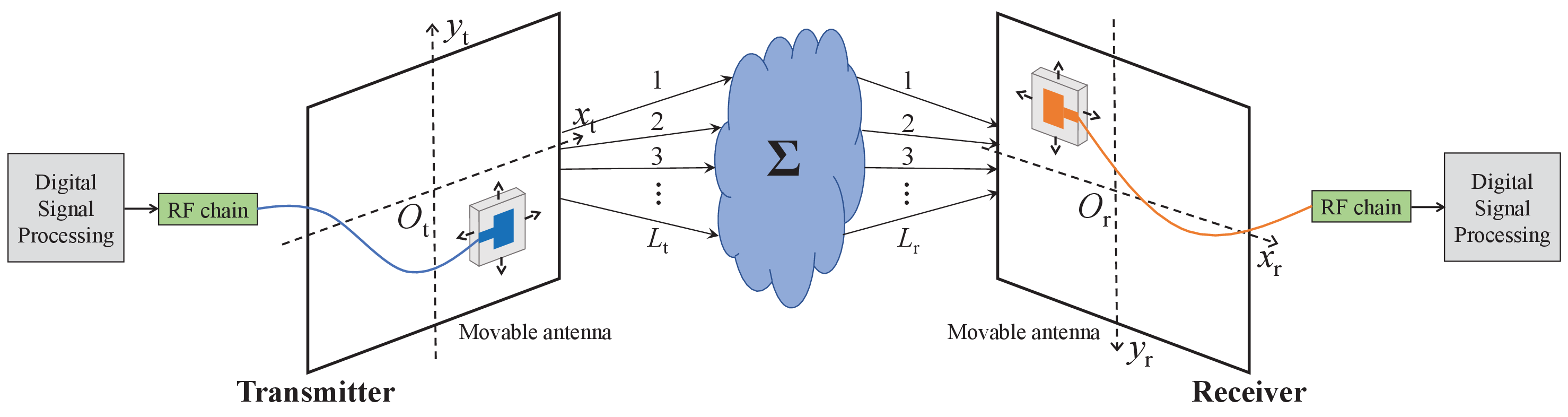}
		\caption{Illustration of the MA-enabled communication system.}
		\label{fig:MovableAntenna}
	\end{center}
\end{figure*}
\begin{figure}[t]
	\begin{center}
		\includegraphics[width=8.8 cm]{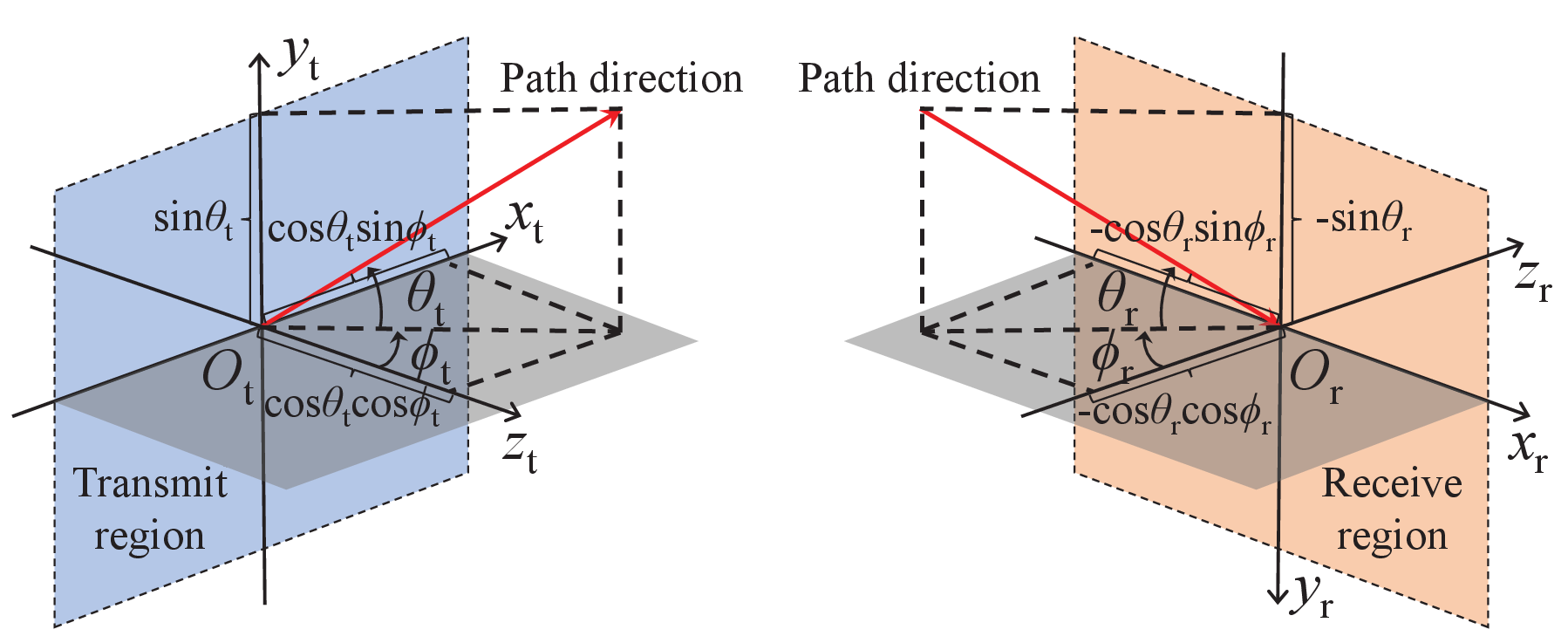}
		\caption{Illustration of the coordinates and spatial angles for transmit and receive regions.}
		\label{fig:Coordinates}
	\end{center}
\end{figure}

The rest of this paper is organized as follows. Section II introduces the signal model and the field-response based channel model for the MA-enabled communication system. In Section III, we show the main analytical results for MA under both deterministic and stochastic channels. Simulation results are provided in Section IV and this paper is concluded in Section V.

\textit{Notation}: $a$, $\mathbf{a}$, $\mathbf{A}$, and $\mathcal{A}$ denote a scalar, a vector, a matrix, and a set, respectively. $(\cdot)^{\rm{T}}$, $(\cdot)^{*}$, and $(\cdot)^{\rm{H}}$ denote transpose, conjugate, and conjugate transpose, respectively. $\mathcal{CN}(\mathbf{0},\mathbf{\Lambda})$ denotes the circularly symmetric complex Gaussian (CSCG) distribution with mean zero and covariance matrix $\mathbf{\Lambda}$. $\mathbb{E}\{\cdot\}$ denotes the expectation of a random variable. $\mathbb{Z}$, $\mathbb{R}$, and $\mathbb{C}$ represent the sets of integer, real, and complex numbers, respectively. $\Re(\cdot)$, $\Im(\cdot)$, $|\cdot|$, and $\angle(\cdot)$ denote the real part, the imaginary part, the amplitude, and the phase of a complex number or complex vector, respectively. $\mathrm{diag}\{\mathbf{a}\}$ is a diagonal matrix with the element in row $i$ and column $i$ equal to the $i$-th element of vector $\mathbf{a}$. $\dif(\cdot)$ and $\partial(\cdot)$ denote the differential and the partial differential of a function, respectively. $\mathbf{1}_{L}$ denotes an $L$-dimensional vector with all the elements equal to 1. $\mathbf{I}_{L}$ denotes an identical matrix of size $L \times L$.

\section{System Model}
The architecture of MA-enabled communication system is shown in Fig. \ref{fig:MovableAntenna}, where the transmit and receive MAs are connected to the RF chains via flexible wires, such as coaxial cables. Thus, the positions of the MAs can be mechanically adjusted with the aid of drive components, such as micro motors. The Cartesian coordinate systems, $x_{\mathrm{t}}$-$O_{\mathrm{t}}$-$y_{\mathrm{t}}$ and $x_{\mathrm{r}}$-$O_{\mathrm{r}}$-$y_{\mathrm{r}}$, are established to describe the positions of the MAs in the transmit and receive regions, $\mathcal{C}_{\mathrm{t}}$ and $\mathcal{C}_{\mathrm{r}}$, respectively\footnote{In this paper, we consider the MAs moving in a 2D plane, which can be extended to the 3D space by adding the $z_{\mathrm{t}}$($z_{\mathrm{r}}$) coordinate. Besides, we assume an omnidirectional MA employed at the Tx/Rx. If a directional MA is used with the capability of rotation, the position and orientation change of the antenna offers six-dimensional (6D) DoFs for MA systems in total. The modeling and performance analysis for directional MA systems are beyond the scope of this paper and will be an interesting topic for future research.}. The coordinates of the transmit MA are denoted as $\mathbf{t}=[x_{\mathrm{t}},y_{\mathrm{t}}]^{\mathrm{T}}$, and the coordinates of the receive MA are denoted as $\mathbf{r}=[x_{\mathrm{r}},y_{\mathrm{r}}]^{\mathrm{T}}$. 

For the considered MA system, the channel response depends on the positions of the antennas. Thus, the channel coefficient can be represented by a function of the positions of the transmit and receive MAs, i.e., $h(\mathbf{t}, \mathbf{r})$. Denoting the transmit power of the Tx as $p_{\mathrm{t}}$, the received signal at the Rx can be expressed as
\begin{equation}\label{eq_signal_model}
	y(\mathbf{t}, \mathbf{r}) = h(\mathbf{t}, \mathbf{r})\sqrt{p_{\mathrm{t}}}s+n,~\mathbf{t} \in \mathcal{C}_{\mathrm{t}}, \mathbf{r} \in \mathcal{C}_{\mathrm{r}},
\end{equation}
where $s$ represents the transmit signal with zero mean and normalized power of one, and $n\sim \mathcal{CN}(0, \delta^{2})$ is the white Gaussian noise at the Rx with power $\delta^{2}$. As a result, the signal-to-noise ratio (SNR) for the received signal is given by
\begin{equation}\label{eq_SNR}
	\gamma(\mathbf{t}, \mathbf{r}) = \frac{\left|h(\mathbf{t}, \mathbf{r})\right|^{2}p_{\mathrm{t}}}{\delta^{2}},~\mathbf{t} \in \mathcal{C}_{\mathrm{t}}, \mathbf{r} \in \mathcal{C}_{\mathrm{r}}.
\end{equation}

\subsection{Field-Response Based Channel Model}
For MA systems, the channel response is determined by the propagation environments and the positions of the antennas. In this paper, we assume that the size of the region for moving the antenna is much smaller than the propagation distance between the Tx and Rx, such that the far-field condition is satisfied at the Tx and Rx. This assumption is reasonable because the typical size of the antenna moving region is in the order of several to tens of wavelengths\footnote{The performance gain provided by antenna movement critically depends on the physical size of the antenna moving region normalized by signal wavelength, as will be shown later.}. For example, in the case of an MA system operating at 30 GHz, the area of a square region with ten-wavelength side length is approximate 0.01 $\mathrm{m}^2$. Denoting $\lambda$ as the signal wavelength, the corresponding Rayleigh distance is obtained as $\frac{2\times (10 \lambda)^{2}}{\lambda}=2$ m \cite{Ning2023THzbeam}, under which the far-field propagation condition can be easily guaranteed in this region for practical Tx-Rx distance. Thus, the plane-wave model can be used to form the field response between the transmit and receive regions. In other words, the AoDs, AoAs, and amplitudes of complex coefficients for the multiple channel paths do not change for different positions of the MAs, while only the phases of the multi-path channels vary in the transmit/receive region.

At the Tx side, we denote the number of transmit paths as $L_{\mathrm{t}}$. As shown in Fig. \ref{fig:Coordinates}, the elevation and azimuth AoDs of the $j$-th transmit path are respectively denoted as $\theta_{\mathrm{t},j} \in [-\pi/2,\pi/2]$ and $\phi_{\mathrm{t},j}\in [-\pi/2,\pi/2]$, $1 \leq j \leq L_{\mathrm{t}}$. At the Rx side, we denote the number of receive paths as $L_{\mathrm{r}}$. The elevation and azimuth AoAs of the $i$-th receive path are respectively denoted as $\theta_{\mathrm{r},i}\in [-\pi/2,\pi/2]$ and $\phi_{\mathrm{r},i}\in [-\pi/2,\pi/2]$, $1 \leq i \leq L_{\mathrm{r}}$. We define a path-response matrix (PRM) $\mathbf{\Sigma} \in \mathbb{C}^{L_{\mathrm{r}} \times L_{\mathrm{t}}}$ to represent the response from the transmit reference position $\mathbf{t}_{0}=(0,0)$ to the receive reference position $\mathbf{r}_{0}=(0,0)$. Specifically, the entry in row $i$ and column $j$ of $\mathbf{\Sigma}$, denoted as $\sigma_{i,j}$, is the response coefficient between the $j$-th transmit path and the $i$-th receive path. As a result, the channel between two antennas located at $\mathbf{t}_{0}$ and $\mathbf{r}_{0}$ is given by
\begin{equation}\label{eq_channel00}
	h(\mathbf{t}_{0}, \mathbf{r}_{0}) = \mathbf{1}_{L_{\mathrm{r}}}^{\mathrm{H}}\mathbf{\Sigma}\mathbf{1}_{L_{\mathrm{t}}},
\end{equation}
which is the linear superposition of all the elements in the PRM.

According to basic geometry shown in Fig. \ref{fig:Coordinates}, the difference of the signal propagation distance between position $\mathbf{r}=[x_{\mathrm{r}},y_{\mathrm{r}}]$ and the reference point $\mathbf{r}_{0}$ is given by $\rho_{\mathrm{r},i}(x_{\mathrm{r}},y_{\mathrm{r}})=x_{\mathrm{r}} \cos \theta_{\mathrm{r},i} \sin \phi_{\mathrm{r},i} + y_{\mathrm{r}} \sin \theta_{\mathrm{r},i}$ for the $i$-th receive path, $1 \leq i \leq L_{\mathrm{r}}$. It indicates that the channel response of the $i$-th receive path at position $\mathbf{r}$ has a $2\pi\rho_{\mathrm{r},i}(x_{\mathrm{r}},y_{\mathrm{r}})/\lambda$ phase difference with respect to the reference point $\mathbf{r}_{0}$, where $\lambda$ is the wavelength. To account for such phase differences in all $L_{\mathrm{r}}$ receive paths, the field-response vector (FRV) in the receive region is defined as
\begin{equation}\label{eq_field_resRx}
	\mathbf{f}(\mathbf{r}) = [e^{j\frac{2\pi}{\lambda}\rho_{\mathrm{r},1}(x_{\mathrm{r}},y_{\mathrm{r}})},
								e^{j\frac{2\pi}{\lambda}\rho_{\mathrm{r},2}(x_{\mathrm{r}},y_{\mathrm{r}})},
								\cdots, e^{j\frac{2\pi}{\lambda}\rho_{\mathrm{r},L_{\mathrm{r}}}(x_{\mathrm{r}},y_{\mathrm{r}})}]^{\mathrm{T}}.
\end{equation}
Similarly, for any position $\mathbf{t}=[x_{\mathrm{t}},y_{\mathrm{t}}]$ in the transmit region, the FRV is defined as
\begin{equation}\label{eq_field_resTx}
	\mathbf{g}(\mathbf{t}) = [e^{j\frac{2\pi}{\lambda}\rho_{\mathrm{t},1}(x_{\mathrm{t}},y_{\mathrm{t}})},
								e^{j\frac{2\pi}{\lambda}\rho_{\mathrm{t},2}(x_{\mathrm{t}},y_{\mathrm{t}})},
								\cdots, e^{j\frac{2\pi}{\lambda}\rho_{\mathrm{t},L_{\mathrm{t}}}(x_{\mathrm{t}},y_{\mathrm{t}})}]^{\mathrm{T}}.
\end{equation}
with $\rho_{\mathrm{t},j}(x_{\mathrm{t}},y_{\mathrm{t}})=x_{\mathrm{t}} \cos \theta_{\mathrm{t},j} \sin \phi_{\mathrm{t},j} + y_{\mathrm{t}} \sin \theta_{\mathrm{t},j}$, $1 \leq j \leq L_{\mathrm{t}}$. As a result, the channel between two antennas located at positions $\mathbf{t}=[x_{\mathrm{t}},y_{\mathrm{t}}]$ and $\mathbf{r}=[x_{\mathrm{r}},y_{\mathrm{r}}]$ is obtained as
\begin{equation}\label{eq_channel_SISO}
	h(\mathbf{t}, \mathbf{r}) = \mathbf{f}(\mathbf{r})^{\mathrm{H}}\mathbf{\Sigma}\mathbf{g}(\mathbf{t}).
\end{equation}

\subsection{Relationship with Conventional Channel Models}
The channel model shown in \eqref{eq_channel_SISO} characterizes the wireless channel based on the transmit/receive antenna locations to facilitate our subsequent performance analysis for MA-enabled communications. For any given positions of the transmit and receive MAs, the channel model in \eqref{eq_channel_SISO} is consistent with that of the conventional FPA systems. In this subsection, we present the conditions under which the model in \eqref{eq_channel_SISO} becomes the well-known LoS channel, geometric channel, Rayleigh and Rician fading channels.
\subsubsection{LoS Channel}
A necessary condition for \eqref{eq_channel_SISO} being equivalent to the LoS channel is that the numbers of transmit and receive paths are both equal to 1, i.e.,
\begin{equation}\label{eq_channel_LoS}
	L_{\mathrm{t}}=L_{\mathrm{r}}=1.
\end{equation}
Under this condition, the PRM and FRVs are all reduced to scalars, i.e.,
\begin{equation}\label{eq_PathRes_LoS}
	\left\{
	\begin{aligned}
	&\mathbf{\Sigma} = \sigma_{1,1},\\
	&\mathbf{f}(\mathbf{r})=e^{j\frac{2\pi}{\lambda}(x_{\mathrm{r}} \cos \theta_{\mathrm{r},1} \sin \phi_{\mathrm{r},1} + y_{\mathrm{r}} \sin \theta_{\mathrm{r},1})},\\
	&\mathbf{g}(\mathbf{t}) = e^{j\frac{2\pi}{\lambda}(x_{\mathrm{t}} \cos \theta_{\mathrm{t},1} \sin \phi_{\mathrm{t},1} + y_{\mathrm{t}} \sin \theta_{\mathrm{t},1})},
	\end{aligned}
	\right.
\end{equation}
where $\sigma_{1,1}$, $\theta_{\mathrm{r},1}$, $\phi_{\mathrm{r},1}$, $\theta_{\mathrm{t},1}$, and $\phi_{\mathrm{t},1}$ represent the path-response coefficient at antenna positions $(\mathbf{t}_{0}, \mathbf{r}_{0})$, the elevation AoA, azimuth AoA, elevation AoD, and azimuth AoD for the LoS path, respectively.
\subsubsection{Geometric Channel}
For geometric channels, the transmit and receive paths have one-to-one correspondence. In other words, each transmit path always arrives at the Rx through one and only one receive path. Thus, a necessary condition for \eqref{eq_channel_SISO} to be the geometric channel is given by
\begin{equation}\label{eq_channel_Geo}
	L_{\mathrm{t}}=L_{\mathrm{r}}=L.
\end{equation}
Under this condition, the PRM becomes diagonal, i.e.,
\begin{equation}\label{eq_PathRes_Geo}
	\mathbf{\Sigma} = \mathrm{diag}\{\sigma_{1,1},\sigma_{2,2},\cdots,\sigma_{L,L} \},
\end{equation}
where $\sigma_{i,i}$, $1 \leq i \leq L$, represents the path-response coefficient corresponding to antenna positions $(\mathbf{t}_{0}, \mathbf{r}_{0})$ for the $i$-th path component between the Tx and Rx.
\subsubsection{Rayleigh Fading Channel}
A necessary condition for \eqref{eq_channel_SISO} to represent Rayleigh fading is that infinite number of statistically independent paths exist between the transmit and receive regions. Thus, the numbers of transmit and receive paths should approach infinity, i.e.,
\begin{equation}\label{eq_channel_Ray}
	L_{\mathrm{t}} \rightarrow +\infty,~L_{\mathrm{r}} \rightarrow +\infty.
\end{equation}
Besides, for isotropic scattering environments, the multi-path components should be uniformly distributed over the half-space in front of the antenna panel, which yields the AoDs and AoAs following the probability density functions (PDFs)
\begin{subequations}\label{eq_angle_Ray}
	\begin{align}
		&f_{\mathrm{AoD}}(\theta_{\mathrm{t}}, \phi_{\mathrm{t}})=\frac{\cos \theta_{\mathrm{t}}}{2\pi},~\theta_{\mathrm{t}} \in \left[-\frac{\pi}{2},\frac{\pi}{2}\right],~\phi_{\mathrm{t}} \in \left[-\frac{\pi}{2},\frac{\pi}{2}\right], \label{eq_angle_Ray_AoD}\\
		&f_{\mathrm{AoA}}(\theta_{\mathrm{r}}, \phi_{\mathrm{r}})=\frac{\cos \theta_{\mathrm{r}}}{2\pi},~\theta_{\mathrm{r}} \in \left[-\frac{\pi}{2},\frac{\pi}{2}\right],~\phi_{\mathrm{r}} \in \left[-\frac{\pi}{2},\frac{\pi}{2}\right], \label{eq_angle_Ray_AoA}
	\end{align}
\end{subequations}
respectively. Under the above conditions, if the elements of the PRM are independent and identically distributed (i.i.d.) circularly symmetric complex random variables, according to the central limit theory, the superimposed path responses at the Tx and Rx, i.e., the elements of $\mathbf{f}(\mathbf{r})^{\mathrm{H}}\mathbf{\Sigma}$ and $\mathbf{\Sigma}\mathbf{g}(\mathbf{t})$, are i.i.d. CSCG distributed random variables. Thus, the channel coefficient $h(\mathbf{t}, \mathbf{r})= \mathbf{f}(\mathbf{r})^{\mathrm{H}}\mathbf{\Sigma}\mathbf{g}(\mathbf{t})$, which is equal to the sum of all the elements of $\mathbf{f}(\mathbf{r})^{\mathrm{H}}\mathbf{\Sigma}$ weighted by the unit-modulus elements of $\mathbf{g}(\mathbf{t})$, or the sum of all the elements of $\mathbf{\Sigma}\mathbf{g}(\mathbf{t})$ weighted by the unit-modulus elements of $\mathbf{f}(\mathbf{r})^{\mathrm{H}}$, is also a CSCG random variable, which leads to Rayleigh fading.

\subsubsection{Rician Fading Channel}
For Rician fading, an LoS path exists between the Tx and the Rx, and the NLoS components are also uniformly distributed over the half-space similar to that of the Rayleigh fading. Thus, the numbers of transmit and receive paths should approach infinity as shown in \eqref{eq_channel_Ray}, and the elements of the PRM are i.i.d. circularly symmetric complex random variables. Meanwhile, the element in row $i^{\star}$ and column $j^{\star}$ of $\mathbf{\Sigma}$ corresponding to the LoS path has a constant amplitude. The Rician factor is thus obtained as
\begin{equation}\label{eq_Rician_fac}
	\kappa = \frac{|\sigma_{i^{\star},j^{\star}}|^2}{\mathbb{E}\left\{\left|\sum \limits_{(i,j) \neq (i^{\star},j^{\star})} \sigma_{i,j}\right|^2\right\}}.
\end{equation}

\section{Performance Analysis}
To analyze the performance gain provided by MAs over conventional FPAs, we consider the simplified scenario where the position of the transmit antenna is fixed at the reference point $\mathbf{t}_{0}$ while the position of the receive antenna can be flexible. Thus, the signal model can be simplified as
\begin{equation}\label{eq_signal_model_single}
	y(\mathbf{r}) = \mathbf{f}(\mathbf{r})^{\mathrm{H}}\mathbf{\Sigma}\mathbf{g}(\mathbf{t}_{0})\sqrt{p_{\mathrm{t}}}s+n \triangleq \mathbf{f}(\mathbf{r})^{\mathrm{H}}\mathbf{b}\sqrt{p_{\mathrm{t}}}s+n,~\mathbf{r} \in \mathcal{C}_{\mathrm{r}},
\end{equation}
where $\mathbf{b}=\mathbf{\Sigma}\mathbf{g}(\mathbf{t}_{0}) \triangleq [b_{1}, b_{2}, \cdots, b_{L_{\mathrm{r}}}]^{\mathrm{T}}$ represents the effective path-response vector (EPRV) in the receive region. Since the channel gain varies with the position of the receive MA, the receive SNR in \eqref{eq_SNR} changes accordingly. In this section, the maximum channel gain achievable by a single receive MA and the corresponding SNR gain over an FPA located at $\mathbf{r}_{0}$ are analyzed under the deterministic and stochastic channel setups, respectively.

\subsection{Deterministic Channel}
In this subsection, we analyze the characteristics of the channel gain in the receive region for deterministic path responses with any given EPRV  $\mathbf{b}$ and physical AoAs $\theta_{\mathrm{r},\ell}$ and $\phi_{\mathrm{r},\ell}$, $1 \leq \ell \leq L_{\mathrm{r}}$. For notation simplicity, we define intermediate variables $\varphi_{\mathrm{r},\ell} = \cos \theta_{\mathrm{r},\ell} \sin \phi_{\mathrm{r},\ell}$ and $\vartheta_{\mathrm{r},\ell} = \sin \theta_{\mathrm{r},\ell}$ as the virtual AoAs for the $\ell$-th receive path, $1 \leq \ell \leq L_{\mathrm{r}}$. Then, the property of the channel gain can be analyzed separately for the one-path, two-path, three-path, and multiple-path cases as follows.
\subsubsection{One-Path Case}
If only one channel path arrives at the Rx, the channel (power) gain is a constant within the receive region, i.e.,
\begin{equation}\label{eq_channel_power1}
	|h_{1}(\mathbf{r})|^{2} = |e^{-j 2\pi(\frac{x_{\mathrm{r}}}{\lambda} \varphi_{\mathrm{r},1}  + \frac{y_{\mathrm{r}}}{\lambda} \vartheta_{\mathrm{r},1})}b_{1}|^{2}=|b_{1}|^{2}.
\end{equation}
In other words, the change of the antenna position can only impact the phase of the channel for the one-path case and the MA cannot provide any SNR gain over the FPA.

\begin{figure}
	\centering
	\includegraphics[width=8 cm]{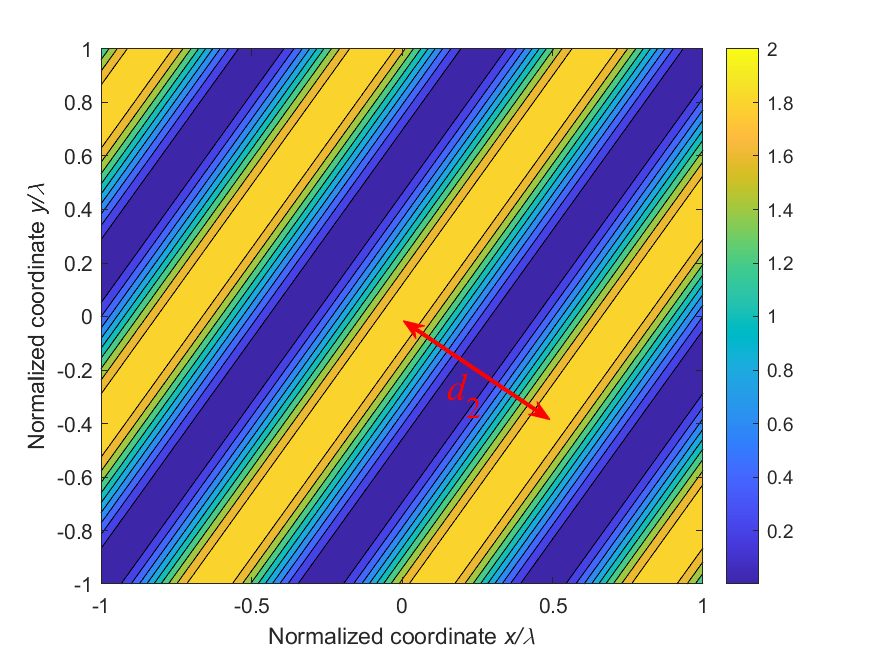}
	\caption{Illustration of the periodic character of the channel gain in the receive region, with $L_{\mathrm{r}}=2$, $b_{1}=b_{2}=\frac{\sqrt{2}}{2}$, $\theta_{\mathrm{r},1}=0$, $\theta_{\mathrm{r},2}=\frac{\pi}{3}$, $\phi_{\mathrm{r},1}=\frac{\pi}{3}$, and $\phi_{\mathrm{r},2}=-\frac{3\pi}{7}$.}
	\label{fig:PowerChannelL2}
\end{figure}

\subsubsection{Two-Path Case}
If two channel paths with different AoAs arrive at the Rx, the channel gain between the transmit antenna and the receive MA at position $\mathbf{r}$ is given by
\begin{equation}\label{eq_channel_power2}
	\begin{aligned}
	|h_{2}(\mathbf{r})|^{2} &= |e^{-j 2\pi(\frac{x_{\mathrm{r}}}{\lambda} \varphi_{\mathrm{r},1}  + \frac{y_{\mathrm{r}}}{\lambda} \vartheta_{\mathrm{r},1})}b_{1}+ e^{-j 2\pi(\frac{x_{\mathrm{r}}}{\lambda} \varphi_{\mathrm{r},2}  + \frac{y_{\mathrm{r}}}{\lambda} \vartheta_{\mathrm{r},2})}b_{2}|^{2}\\
	&=|b_{1}|^{2}+|b_{2}|^{2}+2|b_{1}||b_{2}| \cos \left\{ \omega_{1,2}(x_{\mathrm{r}},y_{\mathrm{r}}) \right\},
	\end{aligned}
\end{equation}
with intermediate variables $\omega_{1,2}(x_{\mathrm{r}},y_{\mathrm{r}})=2\pi \left[\frac{x_{\mathrm{r}}}{\lambda}(\varphi_{\mathrm{r},1}-\varphi_{\mathrm{r},2}) + \frac{y_{\mathrm{r}}}{\lambda} (\vartheta_{\mathrm{r},1}-\vartheta_{\mathrm{r},2})\right]+\mu_{2}-\mu_{1}$ and parameters $\mu_{\ell}=\angle{b_{\ell}}$, $\ell=1,2$. For a rectangular receive region of size $[-A/2, A/2] \times [-A/2, A/2]$, the value range of variable $\omega_{1,2}(x_{\mathrm{r}},y_{\mathrm{r}})$ is from $-\frac{A\pi}{\lambda} \left(|\varphi_{\mathrm{r},1}-\varphi_{\mathrm{r},2}| + |\vartheta_{\mathrm{r},1}-\vartheta_{\mathrm{r},2}|\right)+\mu_{2}-\mu_{1}$ to $\frac{A\pi}{\lambda} \left[|\varphi_{\mathrm{r},1}-\varphi_{\mathrm{r},2}| + |\vartheta_{\mathrm{r},1}-\vartheta_{\mathrm{r},2}|\right]+\mu_{2}-\mu_{1}$. As can be observed, the superimposed power of the two paths has a periodic character in the receive region due to the existence of the cosine function. For any fixed $y_{\mathrm{r}}$, the period of the channel gain along axis $x_{\mathrm{r}}$ is given by $\lambda/(\varphi_{\mathrm{r},1}-\varphi_{\mathrm{r},2})$. It indicates that a larger difference of the virtual AoAs $\varphi_{\mathrm{r},1}$ and $\varphi_{\mathrm{r},2}$ yields a smaller period along axis $x_{\mathrm{r}}$. Similarly, for any fixed $x_{\mathrm{r}}$, the period of the channel gain along axis $y_{\mathrm{r}}$ is given by $\lambda/(\vartheta_{\mathrm{r},1}-\vartheta_{\mathrm{r},2})$. 

To analyze the variation of the channel gain in the receive region, the gradient of the channel gain with respect to the MA position can be derived as
\begin{equation}\label{eq_channel_power2_grad}
	\begin{aligned}
		\nabla |h_{2}(\mathbf{r})|^{2} &= \left[\frac{\partial|h_{2}(\mathbf{r})|^{2}}{\partial x_{\mathrm{r}}}, \frac{\partial|h_{2}(\mathbf{r})|^{2}}{\partial y_{\mathrm{r}}}\right]^{\mathrm{T}}\\
		&=\left[g_{x_{\mathrm{r}},y_{\mathrm{r}}} (\varphi_{\mathrm{r},1}-\varphi_{\mathrm{r},2}),g_{x_{\mathrm{r}},y_{\mathrm{r}}} (\vartheta_{\mathrm{r},1}-\vartheta_{\mathrm{r},2})\right]^{\mathrm{T}},
	\end{aligned}
\end{equation}
with $g_{x_{\mathrm{r}},y_{\mathrm{r}}}=-\frac{4\pi}{\lambda}|b_{1}||b_{2}| \sin \left\{ \omega_{1,2}(x_{\mathrm{r}},y_{\mathrm{r}})\right\}$. Thus, for any position excluding the maximum points, the direction for increasing the channel gain most quickly is given by $\left[(\varphi_{\mathrm{r},1}-\varphi_{\mathrm{r},2}),(\vartheta_{\mathrm{r},1}-\vartheta_{\mathrm{r},2})\right]^{\mathrm{T}}$ or $\left[(\varphi_{\mathrm{r},2}-\varphi_{\mathrm{r},1}),(\vartheta_{\mathrm{r},2}-\vartheta_{\mathrm{r},1})\right]^{\mathrm{T}}$. The MA can be moved along the gradient direction for approaching the maximum channel gain in the receive region most efficiently. Moreover, it can be easily derived that a tight upper bound on the maximum channel gain is
\begin{equation}\label{eq_channel_max2}
	\begin{aligned}
		\max \limits_{\mathbf{r} \in \mathcal{C}_{\mathrm{r}}}|h_{2}(\mathbf{r})|^{2}=(|b_{1}|+|b_{2}|)^{2},
	\end{aligned}
\end{equation}
which can be achieved at positions satisfying
\begin{equation}\label{eq_channel_maxcond2}
	\begin{aligned}
		\left[\frac{x_{\mathrm{r}}}{\lambda}(\varphi_{\mathrm{r},1}-\varphi_{\mathrm{r},2})+\frac{y_{\mathrm{r}}}{\lambda}(\vartheta_{\mathrm{r},1}-\vartheta_{\mathrm{r},2}) \right]+\mu_{2}-\mu_{1}=k,~k \in \mathbb{Z}.
	\end{aligned}
\end{equation}

The equation in \eqref{eq_channel_maxcond2} yields parallel lines in the $x_{\mathrm{r}}$-$y_{\mathrm{r}}$ plane. An example for the channel gain with two receive paths is shown in Fig. \ref{fig:PowerChannelL2}, where the distance of any two adjacent maximum lines can be obtained as
\begin{equation}\label{eq_channel_max_dis2}
	\begin{aligned}
		d_{2}=\frac{\lambda}{\sqrt{(\varphi_{\mathrm{r},1}-\varphi_{\mathrm{r},2})^2+(\vartheta_{\mathrm{r},1}-\vartheta_{\mathrm{r},2})^2}}.
	\end{aligned}
\end{equation}
Note that the tight upper bound shown in \eqref{eq_channel_max2} is not achievable if the size of the receive region is very small. Nonetheless, according to basic geometry, a sufficient condition for achieving the tight upper bound on the channel gain for the two-path case is that the receive region includes a circular area with diameter larger than $d_{2}$. Thus, we know that to achieve the upper bound on the channel gain, a smaller area of the region is required if the AoA difference of the two paths is larger.

\begin{figure}
	\centering
	\includegraphics[width=8 cm]{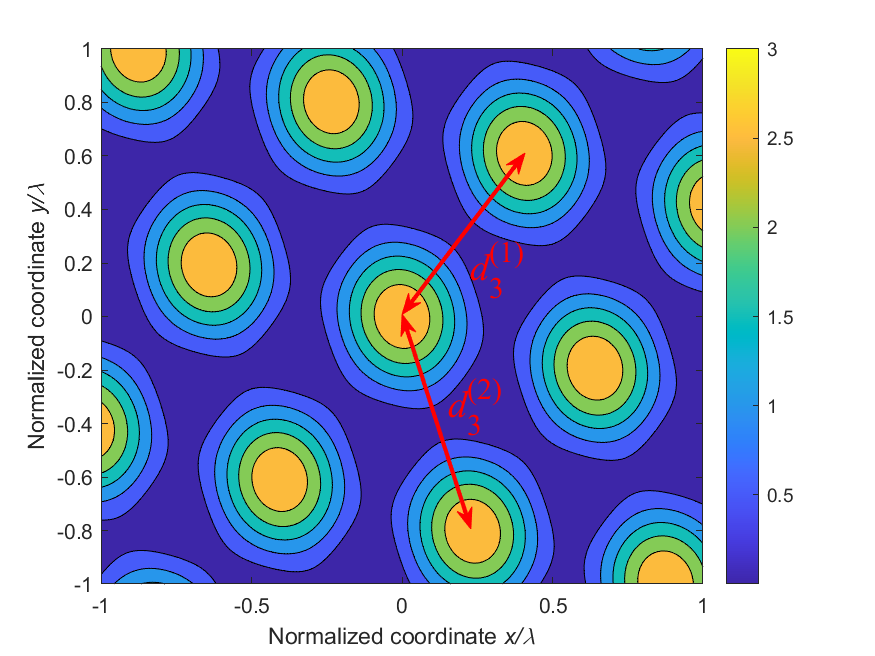}
	\caption{Illustration of the periodic character of the channel gain in the receive region, with $L_{\mathrm{r}}=3$, $b_{1}=b_{2}=b_{3}=\frac{\sqrt{3}}{3}$, $\theta_{\mathrm{r},1}=0$, $\theta_{\mathrm{r},2}=\frac{\pi}{3}$, $\theta_{\mathrm{r},3}=-\frac{\pi}{4}$, $\phi_{\mathrm{r},1}=\frac{\pi}{3}$, $\phi_{\mathrm{r},2}=-\frac{\pi}{3}$, and $\phi_{\mathrm{r},3}=-\frac{3\pi}{7}$.}
	\label{fig:PowerChannelL3}
\end{figure}

\subsubsection{Three-Path Case}
If the number of receive paths is three, the channel gain between the transmit antenna and the receive MA at position $\mathbf{r}$ is given by
\begin{equation}\label{eq_channel_power3}
	\begin{aligned}
		&|h_{3}(\mathbf{r})|^{2} = \left|\sum_{\ell=1}^{3} b_{\ell} e^{-j 2 \pi \left(\frac{x_{\mathrm{r}}}{\lambda} \varphi_{\mathrm{r}, \ell}+\frac{y_{\mathrm{r}}}{\lambda} \vartheta_{\mathrm{r}, \ell}\right)}\right|^{2}\\
		&=|b_{1}|^{2}+|b_{2}|^{2}+|b_{3}|^{2}+2|b_{1}||b_{2}| \cos \left\{ \omega_{1,2}(x_{\mathrm{r}},y_{\mathrm{r}})\right\}\\
		&+2|b_{1}||b_{3}| \cos \left\{ \omega_{1,3}(x_{\mathrm{r}},y_{\mathrm{r}})\right\}+2|b_{2}||b_{3}| \cos \left\{ \omega_{2,3}(x_{\mathrm{r}},y_{\mathrm{r}})\right\},
	\end{aligned}
\end{equation}
with intermediate variables $\omega_{m,n}(x_{\mathrm{r}},y_{\mathrm{r}})=2\pi \left[\frac{x_{\mathrm{r}}}{\lambda}(\varphi_{\mathrm{r},m}-\varphi_{\mathrm{r},n}) + \frac{y_{\mathrm{r}}}{\lambda} (\vartheta_{\mathrm{r},m}-\vartheta_{\mathrm{r},n})\right]+\mu_{n}-\mu_{m}$, $1 \leq m < n \leq 3$, and parameters $\mu_{\ell}=\angle{b_{\ell}}$, $\ell=1,2,3$. For a rectangular receive region of size $[-A/2, A/2] \times [-A/2, A/2]$, the value range of variable $\omega_{m,n}(x_{\mathrm{r}},y_{\mathrm{r}})$ is from $-\frac{A\pi}{\lambda} \left(|\varphi_{\mathrm{r},m}-\varphi_{\mathrm{r},n}| + |\vartheta_{\mathrm{r},m}-\vartheta_{\mathrm{r},n}|\right)+\mu_{n}-\mu_{m}$ to $\frac{A\pi}{\lambda} \left[|\varphi_{\mathrm{r},m}-\varphi_{\mathrm{r},n}| + |\vartheta_{\mathrm{r},m}-\vartheta_{\mathrm{r},n}|\right]+\mu_{n}-\mu_{m}$. It can be observed that the channel gain in \eqref{eq_channel_power3} also has a periodic character. Specifically, let $\mathbf{r}_{\mathrm{p}}=[x_{\mathrm{p}},y_{\mathrm{p}}]^{\mathrm{T}}$ denote the period vector for the channel gain, i.e., we have $|h_{3}(\mathbf{r})|^{2}\equiv|h_{3}(\mathbf{r}+\mathbf{r}_{\mathrm{p}})|^{2}$ for any $\mathbf{r}, \mathbf{r}+\mathbf{r}_{\mathrm{p}} \in \mathcal{C}_{\mathrm{r}}$. A sufficient condition for $\mathbf{r}_{\mathrm{p}}$ being the period of the channel gain is given by
\begin{equation}\label{eq_channel_period3}
	\cos \left\{ \omega_{m,n}(x_{\mathrm{r}},y_{\mathrm{r}}) \right\} \equiv \cos \left\{ \omega_{m,n}(x_{\mathrm{r}}+x_{\mathrm{p}},y_{\mathrm{r}}+y_{\mathrm{p}}) \right\},
\end{equation}
for $1 \leq m < n \leq 3$ and $\forall [x_{\mathrm{r}},y_{\mathrm{r}}]^{\mathrm{T}}, [x_{\mathrm{r}}+x_{\mathrm{p}},y_{\mathrm{r}}+y_{\mathrm{p}}]^{\mathrm{T}} \in \mathcal{C}_{\mathrm{r}}$, which can be equivalently converted to 
\begin{equation}\label{eq_channel_period3eq}
	\frac{x_{\mathrm{p}}}{\lambda}(\varphi_{\mathrm{r},m}-\varphi_{\mathrm{r},n}) + \frac{y_{\mathrm{p}}}{\lambda} (\vartheta_{\mathrm{p},m}-\vartheta_{\mathrm{p},n})=k_{m,n} \in \mathbb{Z}.
\end{equation}
The solution for \eqref{eq_channel_period3eq} can be obtained by directly solving the equation set as
\begin{equation}\label{eq_channel_period3xy}
	\left\{
	\begin{aligned}
		&x_{\mathrm{p}}=\frac{\lambda}{\xi_{1}} \left[k_{1,2} \left(\vartheta_{\mathrm{r}, 1}-\vartheta_{\mathrm{r}, 3}\right)-k_{1,3}\left(\vartheta_{\mathrm{r}, 1}-\vartheta_{\mathrm{r}, 2}\right)\right],\\
		&y_{\mathrm{p}}=\frac{\lambda}{\xi_{2}} \left[k_{1,2} \left(\varphi_{\mathrm{r}, 1}-\varphi_{\mathrm{r}, 3}\right)-k_{1,3} \left(\varphi_{\mathrm{r}, 1}-\varphi_{\mathrm{r}, 2}\right)\right],
	\end{aligned}
	\right.
\end{equation}
with $k_{1,2}\in \mathbb{Z}$, $k_{1,3} \in \mathbb{Z}$, $k_{2,3}=k_{1,3}-k_{1,2}$, $\xi_{1}=\left(\varphi_{\mathrm{r}, 1}-\varphi_{\mathrm{r}, 2}\right)\left(\vartheta_{\mathrm{r}, 1}-\vartheta_{\mathrm{r}, 3}\right)-\left(\varphi_{\mathrm{r}, 1}-\varphi_{\mathrm{r}, 3}\right)\left(\vartheta_{\mathrm{r}, 1}-\vartheta_{\mathrm{r}, 2}\right)$, and $\xi_{2}=\left(\vartheta_{\mathrm{r}, 1}-\vartheta_{\mathrm{r}, 2}\right)\left(\varphi_{\mathrm{r}, 1}-\varphi_{\mathrm{r}, 3}\right)-\left(\vartheta_{\mathrm{r}, 1}-\vartheta_{\mathrm{r}, 3}\right)\left(\varphi_{\mathrm{r}, 1}-\varphi_{\mathrm{r}, 2}\right)$. 

If the arguments in \eqref{eq_channel_period3} are all equal to their maximum value, i.e., one, a tight upper bound on the maximum channel gain can be obtained as
\begin{equation}\label{eq_channel_max3}
	\begin{aligned}
		\max \limits_{\mathbf{r} \in \mathcal{C}_{\mathrm{r}}}|h_{3}(\mathbf{r})|^{2}=(|b_{1}|+|b_{2}|+|b_{3}|)^{2},
	\end{aligned}
\end{equation}
which can be achieved at positions satisfying
\begin{subequations}\label{eq_channel_maxcond3}
	\begin{align}
		&\left[\frac{x_{\mathrm{r}}}{\lambda}(\varphi_{\mathrm{r},1}-\varphi_{\mathrm{r},2})+\frac{y_{\mathrm{r}}}{\lambda}(\vartheta_{\mathrm{r},1}-\vartheta_{\mathrm{r},2})\right]+\mu_{2}-\mu_{1}=k_{1}, \label{eq_channel_maxcond3a}\\
		&\left[\frac{x_{\mathrm{r}}}{\lambda}(\varphi_{\mathrm{r},1}-\varphi_{\mathrm{r},3})+\frac{y_{\mathrm{r}}}{\lambda}(\vartheta_{\mathrm{r},1}-\vartheta_{\mathrm{r},3})\right]+\mu_{3}-\mu_{1}=k_{2}, \label{eq_channel_maxcond3b}\\
		&\left[\frac{x_{\mathrm{r}}}{\lambda}(\varphi_{\mathrm{r},2}-\varphi_{\mathrm{r},3})+\frac{y_{\mathrm{r}}}{\lambda}(\vartheta_{\mathrm{r},2}-\vartheta_{\mathrm{r},3})\right]+\mu_{3}-\mu_{2}=k_{3}, \label{eq_channel_maxcond3c}
	\end{align}
\end{subequations}
with $k_{1},k_{2},k_{3} \in \mathbb{Z} \label{eq_channel_maxcond3a}$. It can be verified that constraint \eqref{eq_channel_maxcond3c} can be safely removed because it holds if and only if constraints \eqref{eq_channel_maxcond3a}, \eqref{eq_channel_maxcond3b}, and $k_{3}=k_{2}-k_{1}$ are satisfied. Thus, the tight upper bound on the maximum channel gain is achieved at the intersections of the lines determined by \eqref{eq_channel_maxcond3a} and \eqref{eq_channel_maxcond3b}, with the coordinates accordingly given by
\begin{equation}\label{eq_channel_maxcoor3}
	\left\{
	\begin{aligned}
		&x_{\mathrm{r}}^{\star}=\frac{\lambda}{2 \pi \xi_{1}} [\left(2 k_{1} \pi+\mu_{1}-\mu_{2}\right)\left(\vartheta_{\mathrm{r}, 1}-\vartheta_{\mathrm{r}, 3}\right)\\
		&~~~~~~~~~~~~~~~~~~-\left(2 k_{2} \pi+\mu_{1}-\mu_{3}\right)\left(\vartheta_{\mathrm{r}, 1}-\vartheta_{\mathrm{r}, 2}\right)],\\
		&y_{\mathrm{r}}^{\star}=\frac{\lambda}{2 \pi \xi_{2}} [\left(2 k_{1} \pi+\mu_{1}-\mu_{2}\right)\left(\varphi_{\mathrm{r}, 1}-\varphi_{\mathrm{r}, 3}\right)\\
		&~~~~~~~~~~~~~~~~~~-\left(2 k_{2} \pi+\mu_{1}-\mu_{3}\right)\left(\varphi_{\mathrm{r}, 1}-\varphi_{\mathrm{r}, 2}\right)],
	\end{aligned}
	\right.
\end{equation}
for $\forall k_{1}, k_{2} \in \mathbb{Z}$. An example for the the channel gain with three receive paths is provided in Fig. \ref{fig:PowerChannelL3}, where the distance of any two adjacent maximum points can be obtained as
\begin{equation}\label{eq_channel_max_dis3}
	\left\{
	\begin{aligned}
		&d_{3}^{(1)}=\lambda \sqrt{\left(\frac{\vartheta_{\mathrm{r}, 1}-\vartheta_{\mathrm{r}, 2}}{\xi_{1}}\right)^{2}+\left(\frac{\varphi_{\mathrm{r}, 1}-\varphi_{\mathrm{r}, 2}}{\xi_{2}}\right)^{2}},\\
		&d_{3}^{(2)}=\lambda \sqrt{\left(\frac{\vartheta_{\mathrm{r}, 1}-\vartheta_{\mathrm{r}, 3}}{\xi_{1}}\right)^{2}+\left(\frac{\varphi_{\mathrm{r}, 1}-\varphi_{\mathrm{r}, 3}}{\xi_{2}}\right)^{2}}.
	\end{aligned}
	\right.
\end{equation}
Similar to the two-path case, the tight upper bound shown in \eqref{eq_channel_max3} is not achievable if the size of the receive region is very small. Nonetheless, according to basic geometry, a sufficient condition for achieving the tight upper bound on the channel gain for the three-path case is that the receive region includes a circular area with diameter larger than $d_{3}=\sqrt{\left(d_{3}^{(1)}\right)^{2}+\left(d_{3}^{(2)}\right)^{2}}$.

\subsubsection{Multiple-Path Case}
If the number of receive paths is larger than three, the channel gain between the transmit antenna and the receive MA at position $\mathbf{r}$ is given by
\begin{equation}\label{eq_channel_powerL}
	\begin{aligned}
		|h_{L_{\mathrm{r}}}(\mathbf{r})|^{2} =\left|\sum_{\ell=1}^{L_{\mathrm{r}}} b_{\ell} e^{-j 2 \pi \left(\frac{x_{\mathrm{r}}}{\lambda} \varphi_{\mathrm{r}, \ell}+\frac{y_{\mathrm{r}}}{\lambda} \vartheta_{\mathrm{r}, \ell}\right)}\right|^{2},
	\end{aligned}
\end{equation}
which has a form similar to the 2D discrete-time Fourier transform (DTFT)\footnote{Note that \eqref{eq_channel_powerL} holds for arbitrary values of $L_{\mathrm{r}} \geq 1$.}. Specifically, the virtual AoAs in the angular domain can be regarded as time domain, while the positions in the spatial domain can be thought as frequency domain. 

\begin{figure}
	\centering
	\includegraphics[width=8 cm]{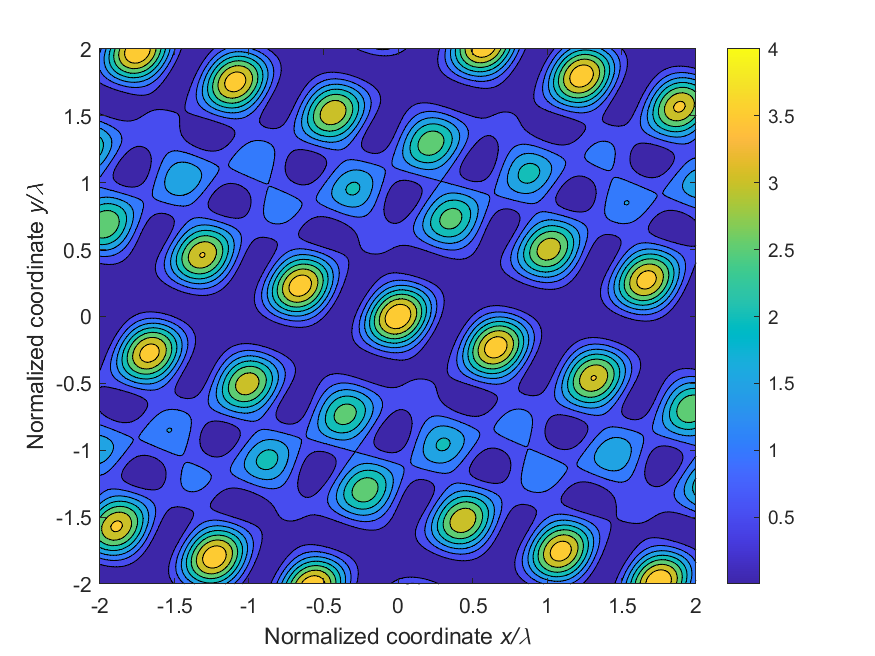}
	\caption{An example for the channel gain in the receive region, with $L_{\mathrm{r}}=4$, $b_{1}=b_{2}=b_{3}=b_{4}=\frac{1}{2}$, $\theta_{\mathrm{r},1}=0$, $\theta_{\mathrm{r},2}=\frac{\pi}{3}$, $\theta_{\mathrm{r},3}=-\frac{\pi}{4}$, $\theta_{\mathrm{r},4}=-\frac{\pi}{3}$, $\phi_{\mathrm{r},1}=\frac{\pi}{3}$, $\phi_{\mathrm{r},2}=-\frac{\pi}{3}$, $\phi_{\mathrm{r},3}=-\frac{3\pi}{7}$, and $\phi_{\mathrm{r},4}=\frac{3\pi}{8}$.}
	\label{fig:PowerChannelL4}
\end{figure}

In general, the period of the channel gain cannot be obtained explicitly for $L_{\mathrm{r}}>3$ as the virtual AoAs $\varphi_{\mathrm{r}, \ell}$ and $\vartheta_{\mathrm{r}, \ell}$ are randomly distributed, which can be observed from Fig. \ref{fig:PowerChannelL4}. Nevertheless, the approximate period for the channel gain can be derived by assuming quantized AoAs. Next, we analyze the periodic character for the channel gain along axis $x_{\mathrm{r}}$, while the periodic character along axis $y_{\mathrm{r}}$ can be derived in a similar way. Specifically, for any fixed $y_{\mathrm{r}}$, we denote the channel gain between the transmit antenna and the receive MA at location $(x_{\mathrm{r}},y_{0})$ as
\begin{equation}\label{eq_channel_powerLx}
	\begin{aligned}
		G(x_{\mathrm{r}})=|h_{L_{\mathrm{r}}}(\mathbf{r})|^{2}_{y_{\mathrm{r}}=y_{0}}
		&=\left|\sum_{\ell=1}^{L_{\mathrm{r}}} b_{\ell} e^{-j 2 \pi \left(\frac{x_{\mathrm{r}}}{\lambda} \varphi_{\mathrm{r}, \ell}+\frac{y_{0}}{\lambda} \vartheta_{\mathrm{r}, \ell}\right)}\right|^{2} \\
		&\triangleq \left|\sum_{\ell=1}^{L_{\mathrm{r}}} \tilde{b}_{\ell} e^{-j 2 \pi \frac{x_{\mathrm{r}}}{\lambda} \varphi_{\mathrm{r}, \ell}}\right|^{2} ,
	\end{aligned}
\end{equation}
with $\tilde{b}_{\ell}=b_{\ell} e^{-j 2 \pi \frac{y_{0}}{\lambda} \vartheta_{\mathrm{r}, \ell}}$. Let $X$ denote the period for the channel gain along axis $x_{\mathrm{r}}$. Then, we have
\begin{equation}\label{eq_channel_power_period}
	\begin{aligned}
		&G(x_{\mathrm{r}}) \equiv G(x_{\mathrm{r}}+X)\\
		\Leftrightarrow &\left|\sum_{\ell=1}^{L_{\mathrm{r}}} \tilde{b}_{\ell} e^{-j 2 \pi \frac{x_{\mathrm{r}}}{\lambda} \varphi_{\mathrm{r}, \ell}}\right|^{2}
		\equiv \left|\sum_{\ell=1}^{L_{\mathrm{r}}} \tilde{b}_{\ell} e^{-j 2 \pi \frac{x_{\mathrm{r}}+X}{\lambda} \varphi_{\mathrm{r}, \ell}}\right|^{2}\\
		\Leftrightarrow &\sum_{m=1}^{L_{\mathrm{r}}} \sum_{n=1}^{L_{\mathrm{r}}} \tilde{b}_{m}^{*} \tilde{b}_{n} e^{j 2 \pi \frac{x_{\mathrm{r}}}{\lambda}\left(\varphi_{\mathrm{r}, m}-\varphi_{\mathrm{r}, n}\right)}\\
		&~~~~~~~~~~~~~~~~~\equiv \sum_{m=1}^{L_{\mathrm{r}}} \sum_{n=1}^{L_{\mathrm{r}}} \tilde{b}_{m}^{*} \tilde{b}_{n} e^{j 2 \pi \frac{x_{\mathrm{r}}+X}{\lambda}\left(\varphi_{\mathrm{r}, m}-\varphi_{\mathrm{r}, n}\right)}\\
		\Leftrightarrow &\sum_{m=1}^{L_{\mathrm{r}}} \sum_{n=1, n \neq m}^{L_{\mathrm{r}}} \tilde{b}_{m}^{*} \tilde{b}_{n}\left[1- e^{-j 2 \pi \frac{X}{\lambda}\left(\varphi_{\mathrm{r}, m}-\varphi_{\mathrm{r}, n}\right)}\right] \\
		&~~~~~~~~~~~~~~~~~~~~~~~~~~~~~\times e^{-j 2 \pi \frac{x_{\mathrm{r}}}{\lambda}\left(\varphi_{\mathrm{r}, m}-\varphi_{\mathrm{r}, n}\right)} \equiv 0\\
		\Leftrightarrow & 1- e^{-j 2 \pi \frac{X}{\lambda}\left(\varphi_{\mathrm{r}, m}-\varphi_{\mathrm{r}, n}\right)}=0, ~1 \leq m,n \leq L_{\mathrm{r}}\\
		\Leftrightarrow & \frac{X}{\lambda}\left(\varphi_{\mathrm{r}, m}-\varphi_{\mathrm{r}, n}\right) \in \mathbb{Z}, ~1 \leq m,n \leq L_{\mathrm{r}},
	\end{aligned}
\end{equation}
which indicates that period $X$ is the minimum real number that ensures $\frac{X}{\lambda}\left(\varphi_{\mathrm{r}, m}-\varphi_{\mathrm{r}, n}\right)$ to be an integer for $1 \leq m,n \leq L_{\mathrm{r}}$. To facilitate performance analysis, we quantize the virtual AoAs with a resolution of $T$, i.e., assuming $\varphi_{\mathrm{r}, \ell} \in \{\Delta_{t}=-1+\frac{2t-1}{T}\}_{1 \leq t \leq T}$. Without loss of generality, we assume that the virtual AoAs are sorted in a non-decreasing order, i.e., $\varphi_{\mathrm{r}, 1} \leq \varphi_{\mathrm{r}, 2} \leq \cdots \leq \varphi_{\mathrm{r}, L_{\mathrm{r}}}$. Besides, we denote the virtual AoAs as $\varphi_{\mathrm{r}, \ell} = \Delta_{t_{\ell}}=-1+\frac{2t_{\ell}-1}{T}$, which means that $\varphi_{\mathrm{r}, \ell}$ corresponds to the $t_{\ell}$-th element in the quantized set of the virtual AoAs. Then, the difference of two adjacent virtual AoAs can be obtained as $\varphi_{\mathrm{r}, \ell+1}-\varphi_{\mathrm{r}, \ell}=\Delta_{t_{\ell+1}}-\Delta_{t_{\ell}}=\frac{2}{T}(t_{\ell+1}-t_{\ell})\triangleq \frac{2\tau_{\ell}}{T}$, $1 \leq \ell \leq L_{\mathrm{r}}$. Denote $\tau^{\star}$ as the maximal common factor for $\{\tau_{\ell}\}_{1 \leq \ell \leq L_{\mathrm{r}}}$, which represents the largest positive integer that divides all the given $\tau_{\ell}$'s without leaving any remainder. To guarantee $\frac{X}{\lambda}\left(\varphi_{\mathrm{r}, m}-\varphi_{\mathrm{r}, n}\right)$ being  integers, the minimum period for the channel gain along axis $x_{\mathrm{r}}$ should be given by
\begin{equation}\label{eq_channel_power_periodL}
	\begin{aligned}
		X=\frac{T\lambda}{2\tau^{\star}}.
	\end{aligned}
\end{equation}

\begin{figure}
	\centering
	\includegraphics[width=8 cm]{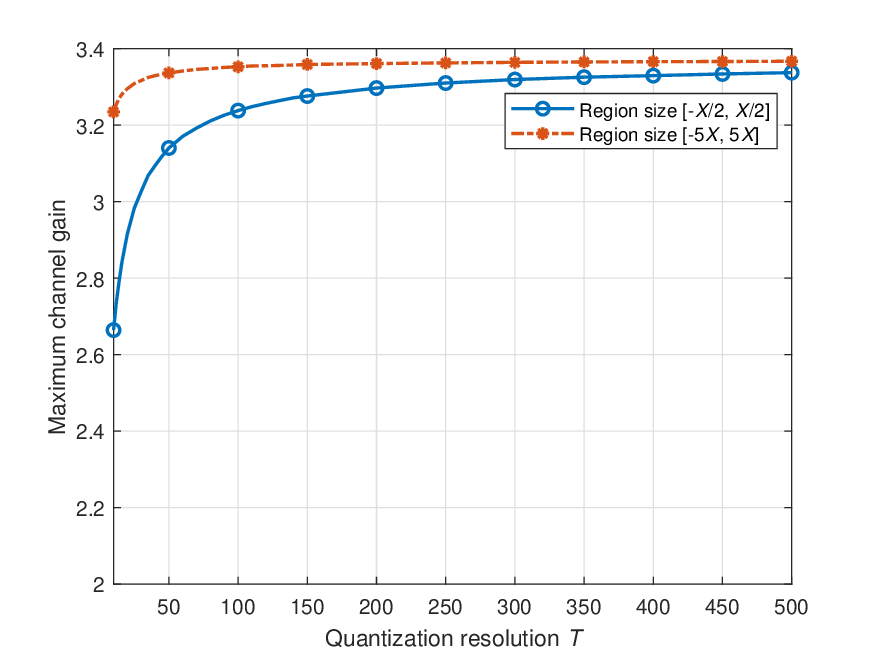}
	\caption{Performance comparison for the maximum channel gains with varying angle quantization resolution, where the number of receive paths is $L_{\mathrm{r}}=4$ and the vertical axis is fixed at $y_{\mathrm{r}}=0$.}
	\label{fig:channel_period}
\end{figure}

Thus, we know that for any fixed $y_{\mathrm{r}}$, the maximum channel gain can always be achieved if the size of the region along axis $x_{\mathrm{r}}$ is no less than period $X$. As can be observed from \eqref{eq_channel_power_periodL}, the accuracy of the period is influenced by the quantization resolution $2/T$ because an approximation on the AoAs is employed. If a small $T$ is used, the difference between the practical AoAs and the approximately quantized AoAs is large, and thus the obtained period becomes inaccurate. In contrast, if a large $T$ is utilized, the obtained period is accurate because the difference between the practical AoAs and the approximately quantized AoAs is small. To evaluate the impact of the quantization resolution, we compare the maximum channel gain for different sizes of the receive region along axis $x_{\mathrm{r}}$. Specifically, the number of receive paths is $L_{\mathrm{r}}=4$, where the virtual AoA $\varphi_{\mathrm{r}, \ell}$ is randomly generated following uniform distribution over $[-1,1]$. The complex path-response coefficient $b_{\ell}$ follows i.i.d. CSCG distribution with mean zero and variance $1/L_{\mathrm{r}}$. For each channel realization, we obtain period $X$ according to \eqref{eq_channel_power_periodL}. Then, the maximum channel gains within regions $x_{\mathrm{r}} \in [-X/2,X/2]$ and $x_{\mathrm{r}} \in [-5X,5X]$ are separately calculated via exhaustive search. Specifically, the receive region is discretized into multiple grids of equal size, $\lambda/100 \times \lambda/100$. Each grid's center serves as the reference point for measuring the channel gain within that grid. Through this process, we identify the grid center that yields the highest channel gain, and this particular location is then chosen as the near-optimal position for the MA. Each point in Fig. \ref{fig:channel_period} is an average result over $10^4$ channel realizations. As can be observed, with the increasing quantization resolution, the performance gap between the two sizes of the region becomes smaller, which indicates that the approximation of the quantizated AoAs has less impact on the maximum value of the channel gain. In other words, a higher quantization resolution can guarantee that the maximum channel gain is more likely achieved within period $X$.

Based on the analysis above, we can observe that the channel gain exhibits a periodic character for the cases of $L_{\mathrm{r}}=2,3$. This indicates that the maximum channel gain can be reaped by moving the antenna within a sub-wavelength region. For the case of $L_{\mathrm{r}} \geq 4$, although the channel gain does not have an explicit period, the more prominent small-scale fading in the spatial domain produces a large number of local maxima, which enable the MAs to improve the channel conditions by small-scale movement in the order of sub-wavelength. For the typical application scenarios experiencing slow channel variation over time, the system may only have finite number of channel patterns shown in Figs. \ref{fig:PowerChannelL2}, \ref{fig:PowerChannelL3}, and \ref{fig:PowerChannelL4} because the AoAs and the amplitudes of the channel paths may remain constant for long periods. As such, the characterization of these channel patterns can provide valuable insights for designing MA-enabled communication systems and optimizing the antenna positions.

\subsection{Stochastic Channel}
Next, we focus on the stochastic performance of the MA system. We consider a stochastic environment where a number of scatters are randomly located around the receive region. Due to the random location of Tx and the random distribution of scatters, it is reasonable to assume that the average power of the channel-path coefficients for different AoAs is identical \cite{goldsmith2005wireless,TseFundaWC}. Thus, the EPRV can be modeled as a CSCG random vector with i.i.d. elements, i.e., $b_{\ell} \sim \mathcal{CN}(0,\sigma^{2}/L_{\mathrm{r}})$, $1\leq \ell \leq L_{\mathrm{r}}$. Besides, the physical AoAs are assumed to be i.i.d. random variables with the PDF given in \eqref{eq_angle_Ray_AoA}. Thus, we can obtain the expected channel gain at the reference point $\mathbf{r}_{0}=(0,0)$ as
\begin{equation}\label{eq_channel_gain0}
	G_{0}=\mathbb{E}\{|h(\mathbf{r}_{0})|^{2}\} = \mathbb{E}\left\{ \left|\sum \limits _{\ell=1}^{L_{\mathrm{r}}}b_{\ell}\right|^{2} \right\}=\sigma^{2}.
\end{equation}
Note that since we assume far-field scenarios, the expected channel gain at any point in the receive region is equal to that at the reference point, i.e., $G(\mathbf{r})=\left|\sum \limits _{\ell=1}^{L_{\mathrm{r}}}e^{-j 2\pi(\frac{x_{\mathrm{r}}}{\lambda} \varphi_{\mathrm{r},\ell}  + \frac{y_{\mathrm{r}}}{\lambda} \vartheta_{\mathrm{r},\ell})}b_{\ell}\right|^{2}=\sigma^{2}$. This equation holds because $b_{\ell}$ is a circularly symmetric random variable for $1\leq \ell \leq L_{\mathrm{r}}$. For the FPA system, the position of the antenna should be fixed at a specific position, and thus the expected channel gain is always equal to $\sigma^{2}$. In contrast, for the MA system, we can always find a position which achieves a larger channel gain in the receive region, and thus an increase on the average channel gain can be acquired over the FPA system.
\subsubsection{Single-Path Case}
As we have analyzed in Section III-A, the channel gain is a constant within the receive region if only one channel path arrives at the Rx. Thus, the maximum channel gain of the MA has an expected value equal to the expected channel gain at the reference point, i.e.,
\begin{equation}\label{eq_channel_stamax1}
	G_{\mathrm{max},1}=G_{0}=\sigma^{2},
\end{equation}
which indicates that no SNR gain is acquired by the MA over the FPA. Furthermore, the CDF of the channel gain for the one-path case can be derived as
\begin{equation}\label{eq_channel_staCDF1}
	\begin{aligned}
		F_{1}(t) \triangleq \Pr \left\{|b_{1}|^2 \leq t\right\}
		=\int_{0}^{\sqrt{t}} \frac{2x}{\sigma^{2}}e^{-\frac{x^{2}}{\sigma^{2}}} \dif x
		=1-e^{-\frac{t}{\sigma^{2}}},
	\end{aligned}
\end{equation}
for $t \geq 0$. For any given transmit power $p_{\mathrm{t}}$, noise power $\delta^{2}$, and receive SNR threshold $\gamma_{\mathrm{th}}$, the outage probability for the MA system with a single receive path is given by $F_{1}(\frac{\delta^{2}\gamma_{\mathrm{th}}}{p_{\mathrm{t}}})$.
\subsubsection{Two-Path Case}
According to Section III-A, for any given $\varphi_{\mathrm{r},1}, \varphi_{\mathrm{r},2}, \vartheta_{\mathrm{r},1}, \vartheta_{\mathrm{r},2}$, the maximum channel gain $(|b_{1}|+|b_{2}|)^2$ can always be achieved if the diameter of the receive region is larger than $d_{2}$ shown in \eqref{eq_channel_max_dis2}. Since $b_{\ell}$ is a CSCG random variable with mean zero and variance $\sigma^{2}/L_{\mathrm{r}}$, $|b_{\ell}|$ is a Rayleigh-distributed random variable with the PDF given by
\begin{equation}\label{eq_path_PDF2}
	f_{|b_{\ell}|}(t)=\frac{4t}{\sigma^{2}}e^{\frac{-2t^{2}}{\sigma^{2}}},~t \geq 0,~\ell=1,2,
\end{equation}

Thus, as the size of the receive region approaches infinity, the upper bound on the maximum channel gain can always be achieved for arbitrary AoAs. The expected value of the maximum channel gain of the MA in the receive region is thus given by
\begin{equation}\label{eq_channel_stamax2}
	\begin{aligned}
	G_{\mathrm{max},2}&=\mathbb{E}\left\{ (|b_{1}|+|b_{2}|)^2 \right\}\\
	&=\int_{0}^{+\infty}\int_{0}^{+\infty}(x+y)^2 \times \frac{4x}{\sigma^{2}}e^{\frac{-2x^{2}}{\sigma^{2}}} \times \frac{4y}{\sigma^{2}}e^{\frac{-2y^{2}}{\sigma^{2}}} \dif x \dif y\\
	&=2\int_{0}^{+\infty}\frac{4x^{3}}{\sigma^{2}}e^{-\frac{2x^{2}}{\sigma^{2}}}\dif x \times 
	 \int_{0}^{+\infty}\frac{4y}{\sigma^{2}}e^{-\frac{2y^{2}}{\sigma^{2}}}\dif y\\
	&+2\int_{0}^{+\infty}\frac{4x^{2}}{\sigma^{2}}e^{-\frac{2x^{2}}{\sigma^{2}}}\dif x \times
	 \int_{0}^{+\infty}\frac{4y^{2}}{\sigma^{2}}e^{-\frac{2y^{2}}{\sigma^{2}}}\dif y\\
	&=2\times \frac{\sigma^{2}}{2}+\frac{\pi\sigma^{2}}{4}=\left(1+\frac{\pi}{4}\right)\sigma^{2},
	\end{aligned}
\end{equation}
with the values of the definite integrals $\int_{0}^{+\infty}\frac{4x}{\sigma^{2}}e^{-\frac{2x^{2}}{\sigma^{2}}}\dif x=1$, $\int_{0}^{+\infty}\frac{4x^{2}}{\sigma^{2}}e^{-\frac{2x^{2}}{\sigma^{2}}}\dif x= \frac{\sigma}{2}\sqrt{\frac{\pi}{2}}$, and $\int_{0}^{+\infty}\frac{4x^{3}}{\sigma^{2}}e^{-\frac{2x^{2}}{\sigma^{2}}}\dif x= \frac{\sigma^{2}}{2}$. Thus, for the two-path case, the average-SNR gain of the MA system over the FPA system is obtained as
\begin{equation}\label{eq_SNR_gain2}
	\begin{aligned}
		\eta_{2}=\frac{\max \limits_{\mathbf{r} \in \mathcal{C}_{\mathrm{r}}}\gamma_{2}(\mathbf{r})}{\gamma_{2}(\mathbf{r}_{0})}=\frac{G_{\mathrm{max},2}}{G_{0}}=1+\frac{\pi}{4}.
	\end{aligned}
\end{equation}

Furthermore, the CDF of the maximum channel gain for the two-path case can be derived as
\begin{equation}\label{eq_channel_staCDF2}
	\begin{aligned}
		F_{2}(t) &\triangleq \Pr \left\{(|b_{1}|+|b_{2}|)^2 \leq t\right\}\\
		&=\int_{0}^{\sqrt{t}} \dif x \int_{0}^{\sqrt{t}-x} \frac{4x}{\sigma^{2}}e^{-\frac{2x^{2}}{\sigma^{2}}} \times \frac{4y}{\sigma^{2}}e^{-\frac{2y^{2}}{\sigma^{2}}} \dif y\\
		&=\int_{0}^{\sqrt{t}} \frac{4x}{\sigma^{2}}e^{-\frac{2x^{2}}{\sigma^{2}}} \times \left[1-e^{-\frac{2(\sqrt{t}-x)^{2}}{\sigma^{2}}}\right] \dif x\\
		&=1-e^{-\frac{2t}{\sigma^{2}}}-e^{-\frac{t}{\sigma^{2}}} \int_{0}^{\sqrt{t}} \frac{4x}{\sigma^{2}} e^{-\frac{(2x-\sqrt{t})^{2}}{\sigma^{2}}}  \dif x \\
		&=1-e^{-\frac{2t}{\sigma^{2}}}-\frac{2\sqrt{\pi t}}{\sigma} e^{-\frac{t}{\sigma^{2}}}  \int_{0}^{\sqrt{2t}} \frac{1}{\sqrt{2\pi}\sigma} e^{-\frac{x^{2}}{2\sigma^{2}}}  \dif x\\
		&=1-e^{-\frac{2t}{\sigma^{2}}}-\frac{\sqrt{\pi t}}{\sigma} e^{-\frac{t}{\sigma^{2}}}  \left[1-2Q\left(\frac{\sqrt{2t}}{\sigma}\right)\right],~t \geq 0.
	\end{aligned}
\end{equation}
with $Q(t)=\int_{t}^{+ \infty} \frac{1}{\sqrt{2\pi}} e^{-\frac{x^{2}}{2}}  \dif x$ being the Gaussian tail probability. For any given transmit power $p_{\mathrm{t}}$, noise power $\delta^{2}$, and receive SNR threshold $\gamma_{\mathrm{th}}$, the outage probability for the MA system with two receive paths is given by $F_{2}(\frac{\delta^{2}\gamma_{\mathrm{th}}}{p_{\mathrm{t}}})$.

\subsubsection{Three-Path Case}
For any given $\varphi_{\mathrm{r},1}, \varphi_{\mathrm{r},2}, \varphi_{\mathrm{r},3}, \vartheta_{\mathrm{r},1}, \vartheta_{\mathrm{r},2}, \vartheta_{\mathrm{r},3}$, the maximum channel gain $(|b_{1}|+|b_{2}|+|b_{3}|)^2$ can always be achieved if the diameter of the receive region is larger than $d_{3}$ as shown in \eqref{eq_channel_max_dis3}. Since $b_{\ell}$ is a CSCG random variable with mean zero and variance $\sigma^{2}/L_{\mathrm{r}}$, $|b_{\ell}|$ is Rayleigh distributed with the PDF given by
\begin{equation}\label{eq_path_PDF3}
	f_{|b_{\ell}|}(t)=\frac{6t}{\sigma^{2}}e^{\frac{-3t^{2}}{\sigma^{2}}},~t \geq 0,~\ell=1,2,3,
\end{equation}

Thus, as the size of the receive region approaches infinity, the upper bound on the maximum channel gain can always be achieved for arbitrary AoAs. The expected value of the maximum channel gain of the MA in the receive region is thus given by
\begin{equation}\label{eq_channel_stamax3}
	\begin{aligned}
		G_{\mathrm{max},3}&=\mathbb{E}\left\{ (|b_{1}|+|b_{2}|+|b_{3}|)^2 \right\}\\
		&=\int_{0}^{+\infty}\int_{0}^{+\infty}\int_{0}^{+\infty}(x+y+z)^2 \times \frac{6x}{\sigma^{2}}e^{\frac{-3x^{2}}{\sigma^{2}}} \\ 
		&~~~~~~~~~~~~~~~~~\times \frac{6y}{\sigma^{2}}e^{\frac{-3y^{2}}{\sigma^{2}}} \times \frac{6z}{\sigma^{2}}e^{\frac{-3z^{2}}{\sigma^{2}}} \dif x \dif y \dif z\\
		&=3\int_{0}^{+\infty}\frac{6x^{3}}{\sigma^{2}}e^{-\frac{3x^{2}}{\sigma^{2}}}\dif x \times 
		\int_{0}^{+\infty}\frac{6y}{\sigma^{2}}e^{-\frac{3y^{2}}{\sigma^{2}}}\dif y\\ 
		&~\times \int_{0}^{+\infty}\frac{6z}{\sigma^{2}}e^{-\frac{3z^{2}}{\sigma^{2}}}\dif z
		+6\int_{0}^{+\infty}\frac{6x^{2}}{\sigma^{2}}e^{-\frac{3x^{2}}{\sigma^{2}}}\dif x \\
		&~\times \int_{0}^{+\infty}\frac{6y^{2}}{\sigma^{2}}e^{-\frac{3y^{2}}{\sigma^{2}}}\dif y \times
		\int_{0}^{+\infty}\frac{6z}{\sigma^{2}}e^{-\frac{3z^{2}}{\sigma^{2}}}\dif z\\
		&=3 \times \frac{\sigma^{2}}{3} +6 \times \frac{\pi\sigma^{2}}{12}\\
		& =\left(1+\frac{\pi}{2}\right)\sigma^{2},
	\end{aligned}
\end{equation}
with the values of the definite integrals $\int_{0}^{+\infty}\frac{6x}{\sigma^{2}}e^{-\frac{3x^{2}}{\sigma^{2}}}\dif x=1$, $\int_{0}^{+\infty}\frac{6x^{2}}{\sigma^{2}}e^{-\frac{3x^{2}}{\sigma^{2}}}\dif x= \frac{\sigma}{\sqrt{6}}\sqrt{\frac{\pi}{2}}$, and $\int_{0}^{+\infty}\frac{6x^{3}}{\sigma^{2}}e^{-\frac{3x^{2}}{\sigma^{2}}}\dif x= \frac{\sigma^{2}}{3}$. Thus, for the three-path case, the average-SNR gain of the MA system over the FPA system is obtained as
\begin{equation}\label{eq_SNR_gain3}
	\begin{aligned}
		\eta_{3}=\frac{\max \limits_{\mathbf{r} \in \mathcal{C}_{\mathrm{r}}}\gamma_{3}(\mathbf{r})}{\gamma_{3}(\mathbf{r}_{0})}=\frac{G_{\mathrm{max},3}}{G_{0}}=1+\frac{\pi}{2}.
	\end{aligned}
\end{equation}

Note that the maximum channel gain is the square-sum of three i.i.d. Rayleigh random variables, which has no closed-form CDF. Nevertheless, the CDF of the maximum channel gain for the three-path case can be approximated as \cite{Hu2005Accura}
\begin{equation}\label{eq_channel_staCDF3}
	\begin{aligned}
		F_{3}(t) &\triangleq \Pr \left\{\left(|b_{1}|+|b_{2}+|b_{3}|\right)^2 \leq t\right\}\\
		&=\Pr \left\{|b_{1}|+|b_{2}+|b_{3}| \leq \sqrt{t}\right\}\\
		&\approx 1-e^{-\frac{t}{c_{3}}}\left(1+\frac{t}{c_{3}}+\frac{t^{2}}{2c_{3}^{2}}\right),~t \geq 0,
	\end{aligned}
\end{equation}
with $c_{3}=\frac{15^{1/3}\sigma^{2}}{L_{\mathrm{r}}}$. For any given transmit power $p_{\mathrm{t}}$, noise power $\delta^{2}$, and receive SNR threshold $\gamma_{\mathrm{th}}$, the outage probability for the MA system with three receive paths is approximately given by $F_{3}(\frac{\delta^{2}\gamma_{\mathrm{th}}}{p_{\mathrm{t}}})$.

\subsubsection{Multiple-Path Case}
Since the EPRV $\mathbf{b}$ is a CSCG random vector following distribution $\mathbf{b} \sim \mathcal{CN}(\mathbf{0},\frac{\sigma^{2}}{L_{\mathrm{r}}}\mathbf{I}_{L_{\mathrm{r}}})$, the amplitude of the $\ell$-th element of $\mathbf{b}$, i.e., $|b_{\ell}|$, is i.i.d. Rayleigh distributed with the PDF given by
\begin{equation}\label{eq_path_PDFL}
	f_{|b_{\ell}|}(t)=\frac{2L_{\mathrm{r}}t}{\sigma^{2}}e^{\frac{-L_{\mathrm{r}}t^{2}}{\sigma^{2}}},~t \geq 0,~1 \leq \ell \leq L_{\mathrm{r}},
\end{equation}
If the number of receive paths is larger than three, it is difficult to derive the explicit expression of the maximum channel gain. Nevertheless, we can always obtain an upper bound on the maximum channel gain of the MA as follows:
\begin{equation}\label{eq_channel_stamaxL_bound}
	\begin{aligned}
		&G_{\mathrm{max},L_{\mathrm{r}}} \leq \mathbb{E}\left\{ \left(\sum \limits _{\ell=1}^{L_{\mathrm{r}}} |b_{\ell}|\right)^{2} \right\}\\
		&=\idotsint_{0}^{+\infty} \left(\sum \limits _{\ell=1}^{L_{\mathrm{r}}} x_{\ell}\right)^2 \times 
		\prod \limits _{\ell=1}^{L_{\mathrm{r}}} \frac{2L_{\mathrm{r}}x_{\ell}}{\sigma^{2}}e^{\frac{-L_{\mathrm{r}}x_{\ell}^{2}}{\sigma^{2}}} \dif x_{1} \cdots \dif x_{L_{\mathrm{r}}}\\
		&=L_{\mathrm{r}}\int_{0}^{+\infty}\frac{2L_{\mathrm{r}}x^{3}}{\sigma^{2}}e^{-\frac{L_{\mathrm{r}}x^{2}}{\sigma^{2}}}\dif x 
		+L_{\mathrm{r}}(L_{\mathrm{r}}-1)\\
		&~~\times \int_{0}^{+\infty}\frac{2L_{\mathrm{r}}x}{\sigma^{2}}e^{-\frac{L_{\mathrm{r}}x^{2}}{\sigma^{2}}}\dif x \times 
		\int_{0}^{+\infty}\frac{2L_{\mathrm{r}}y^{3}}{\sigma^{2}}e^{-\frac{L_{\mathrm{r}}y^{2}}{\sigma^{2}}}\dif y \\
		&=\frac{\sigma^{2}}{L_{\mathrm{r}}} \times L_{\mathrm{r}}+\frac{\pi\sigma^{2}}{4L_{\mathrm{r}}} \times L_{\mathrm{r}}(L_{\mathrm{r}}-1)\\
		&=\left[1+\frac{(L_{\mathrm{r}}-1)\pi}{4}\right]\sigma^{2},
	\end{aligned}
\end{equation}
with the values of the definite integrals $\int_{0}^{+\infty}\frac{2L_{\mathrm{r}}x}{\sigma^{2}}e^{-\frac{L_{\mathrm{r}}x^{2}}{\sigma^{2}}}\dif x=1$, $\int_{0}^{+\infty}\frac{2L_{\mathrm{r}}x^{2}}{\sigma^{2}}e^{-\frac{L_{\mathrm{r}}x^{2}}{\sigma^{2}}}\dif x= \frac{\sigma}{\sqrt{2L_{\mathrm{r}}}}\sqrt{\frac{\pi}{2}}$, and $\int_{0}^{+\infty}\frac{2L_{\mathrm{r}}x^{3}}{\sigma^{2}}e^{-\frac{L_{\mathrm{r}}x^{2}}{\sigma^{2}}}\dif x= \frac{\sigma^{2}}{L_{\mathrm{r}}}$. Hence, for the multiple-path case, the average-SNR gain of the MA system over the FPA system is upper-bounded by
\begin{equation}\label{eq_SNR_gainL}
	\begin{aligned}
		\eta_{L_{\mathrm{r}}}=\frac{\max \limits_{\mathbf{r} \in \mathcal{C}_{\mathrm{r}}}\gamma_{L_{\mathrm{r}}}(\mathbf{r})}{\gamma_{L_{\mathrm{r}}}(\mathbf{r}_{0})}
		=\frac{G_{\mathrm{max},L_{\mathrm{r}}}}{G_{0}} \leq 1+\frac{(L_{\mathrm{r}}-1)\pi}{4}.
	\end{aligned}
\end{equation}

Note that the CDF of the square-sum of $L_{\mathrm{r}}$ independent Rayleigh random variables has no closed-form expression for $L_{\mathrm{r}} > 3$. Nevertheless, the CDF of the upper bound on the maximum channel gain for the multiple-path case can be obtained by the small argument approximation as follows \cite{Hu2005Accura}:
\begin{equation}\label{eq_channel_staCDFL}
	\begin{aligned}
		F_{L_{\mathrm{r}}}^{\mathrm{UB}}(t) &\triangleq \Pr \left\{\left(\sum \limits _{\ell=1}^{L_{\mathrm{r}}} |b_{\ell}|\right)^{2} \leq t\right\} \\ 
		&=\Pr \left\{\sum \limits _{\ell=1}^{L_{\mathrm{r}}} |b_{\ell}| \leq \sqrt{t}\right\}\\
		&\approx 1-e^{-\frac{t}{c}} \sum_{k=0}^{L_{\mathrm{r}}-1} \frac{\left(\frac{t}{c}\right)^{k}}{k !},~t \geq 0,
	\end{aligned}
\end{equation}
with $c=\frac{\sigma^{2}}{L_{\mathrm{r}}}[(2 L_{\mathrm{r}}-1) ! !]^{1 / L_{\mathrm{r}}}$ and $(2 L_{\mathrm{r}}-1) ! !=(2 L_{\mathrm{r}}-1)\times(2 L_{\mathrm{r}}-3)\times \cdots \times 3 \times 1$. For any given transmit power $p_{\mathrm{t}}$, noise power $\delta^{2}$, and receive SNR threshold $\gamma_{\mathrm{th}}$, a lower bound on the outage probability of the MA system with $L_{\mathrm{r}}>3$ receive paths is approximately given by $F_{L_{\mathrm{r}}}^{\mathrm{UB}}(\frac{\delta^{2}\gamma_{\mathrm{th}}}{p_{\mathrm{t}}})$.

\subsubsection{Infinite-Path Case}
For the isotropic scattering scenario, the number of receive paths approaches infinity with the joint PDF for the physical AoAs given by \eqref{eq_angle_Ray_AoA}. According to the central limit theorem, the channel coefficient at each position is a CSCG random variable with mean zero and variance $\sigma^{2}$. Let $b(\theta_{\mathrm{r}},\phi_{\mathrm{r}})$ denote the complex path-response coefficient corresponding to AoAs $\theta_{\mathrm{r}}$ and $\phi_{\mathrm{r}}$, which has an average power of $\sigma^{2}$. The correlation between the channel coefficients of two positions spaced by $d$ can be derived as\footnote{Without loss of generality, the two positions spaced by $d$ are selected along axis $y_{\mathrm{r}}$. For any other two locations in the receive region with spacing $d$, we can always establish a new Cartesian coordinate system with the two points located on axis $y_{\mathrm{r}}$, and the distribution of the corresponding AoAs does not change due to the assumption of isotropic scattering.}
\begin{equation}\label{eq_channel_cov}
	\begin{aligned}
		&R(d)=\mathbb{E}\left\{h^{*}(0,d)h(0,0)\right\}\\
		&=\mathbb{E}\left\{\int_{-\pi/2}^{\pi/2}\int_{-\pi/2}^{\pi/2} |b(\theta_{\mathrm{r}},\phi_{\mathrm{r}})|^{2} e^{j2\pi\frac{d}{\lambda}\sin\theta_{\mathrm{r}}} \frac{\cos\theta_{\mathrm{r}}}{2\pi} \dif\theta_{\mathrm{r}} \dif\phi_{\mathrm{r}} \right\}\\
		&=\int_{-\pi/2}^{\pi/2} \sigma^{2}e^{j2\pi\frac{d}{\lambda}\sin\theta_{\mathrm{r}}} \frac{\cos\theta_{\mathrm{r}}}{2} \dif\theta_{\mathrm{r}}
		=\int_{-1}^{1} \frac{\sigma^{2}}{2}e^{j2\pi\frac{d}{\lambda}t} \dif t \\
		&= \sigma^{2}\frac{\sin\left(2\pi\frac{d}{\lambda}\right)}{2\pi\frac{d}{\lambda}} = \sigma^{2}\mathrm{sinc}\left(\frac{2d}{\lambda}\right).
	\end{aligned}
\end{equation}
According to the property of the $\mathrm{sinc}$ function, we know that the channels of two positions spaced by $\lambda/2$ are statistically independent. Due to the small value of $\mathrm{sinc}(t)$ for $t \geq 1$, we can further assume that the channels of two positions with distance larger than $\lambda/2$ are statistically independent. Thus, for a square receive region with size $A \times A$, the number of independent random channel variables is no smaller than $N_{\mathrm{LB}}=\lfloor 2A/\lambda+1 \rfloor ^{2}$. Since the channel coefficients are CSCG random variables, the channel (power) gains are exponential random variables (ERVs) with the rate parameter $1/\sigma^{2}$. The maximum channel gain of the MA in the receive region is lower-bounded by the largest order statistics of the $N_{\mathrm{LB}}$ independent ERVs, which can be expressed as $X_{\max}=\max \limits _{1 \leq k \leq N_{\mathrm{LB}}} X_{k}$ with $\left\{X_{k}\right\}$ being i.i.d. ERVs with the rate parameter $1/\sigma^{2}$. It is known that the largest order statistics for i.i.d. ERVs can be equivalently expressed as $X_{\max}=\sum \limits _{k=1}^{N_{\mathrm{LB}}} Y_{k}$, where $\left\{Y_{k}\right\}$ are independent ERVs with the rate parameter $k/\sigma^{2}$ for $Y_{k}$, $1 \leq k \leq N_{\mathrm{LB}}$ \cite{rohatgi2015introduction}. Thus, the expected value for the lower bound on the maximum channel gain in the $A \times A$ area is obtained as
\begin{equation}\label{eq_channel_stamaxInfLB}
	\begin{aligned}
		G_{\mathrm{max},\infty} \geq \mathbb{E} \left\{\max \limits_{1 \leq k \leq N_{\mathrm{LB}}} X_{k}\right\}= \mathbb{E} \left\{\sum \limits _{k=1}^{N_{\mathrm{LB}}} Y_{k}\right\}= \sum \limits_{k=1}^{N_{\mathrm{LB}}}\frac{\sigma^{2}}{k},
	\end{aligned}
\end{equation}
which monotonously increases with the size of the receive region and approaches infinity for an infinite region. Besides, the CDF of the lower bound on the maximum channel gain for the infinite-path case is given by
\begin{equation}\label{eq_channel_staCDFInfLB}
	\begin{aligned}
		F_{\infty}^{\mathrm{LB}}(t) = \Pr \left\{\max \limits_{1 \leq k \leq N_{\mathrm{LB}}} X_{k} \leq t \right\}= \left(1-e^{-\frac{t}{\sigma^{2}}}\right)^{N_{\mathrm{LB}}}.
	\end{aligned}
\end{equation}
Thus, for any given transmit power $p_{\mathrm{t}}$, noise power $\delta^{2}$, and receive SNR threshold $\gamma_{\mathrm{th}}$, an upper bound on the outage probability for the MA system with infinite number of receive paths is given by $F_{\infty}^{\mathrm{LB}}(\frac{\delta^{2}\gamma_{\mathrm{th}}}{p_{\mathrm{t}}})$.

To analyze the upper bound on the maximum channel gain of the MA, we discretize the square receive region $A \times A$ into equally spaced grids with size $1/P \times 1/P$, and thus the total number of the grids is no larger than $N_{\mathrm{UB}}=\lceil  PA+1 \rceil ^{2}$. Let $\mathcal{C}_{\mathrm{r},k}$ denote the $k$-th grid in the receive region, $1 \leq k \leq N_{\mathrm{UB}}$. Thus, the maximum channel gain of the MA in the $A \times A$ area can be equivalently expressed as
\begin{equation}\label{eq_channel_maxdis}
	\begin{aligned}
		\max \limits_{\mathbf{r}\in \mathcal{C}_{\mathrm{r}}} |h(\mathbf{r})|^{2} = \max \limits_{1 \leq k \leq N} \max \limits_{\mathbf{r}\in \mathcal{C}_{\mathrm{r},k}}|h(\mathbf{r})|^{2}.
	\end{aligned}
\end{equation}
For sufficiently large $P$, the area of each grid approaches zero such that the correlation between channels at any two points within the same grid approaches $\sigma^{2}$. It indicates that the maximum channel gain over each grid can be approximately given by the channel gain at the center of this grid, i.e., $\max \limits_{\mathbf{r}\in \mathcal{C}_{\mathrm{r},k}}|h(\mathbf{r})|^{2} \approxeq \tilde{X}_{k}$ with $\{\tilde{X}_{k}\}$ being i.i.d. exponential variables with the rate parameter $1/\sigma^{2}$. Thus, an upper bound on the maximum channel gain can be obtained by assuming the channel gains at the center of the $N_{\mathrm{UB}}$ grids are all statistically independent, i.e.,
\begin{equation}\label{eq_channel_stamaxInfUB}
	\begin{aligned}
		G_{\mathrm{max},\infty} \leq \mathbb{E} \left\{\max \limits_{1 \leq k \leq N_{\mathrm{UB}}} \tilde{X}_{k}\right\}= \sum \limits_{k=1}^{N_{\mathrm{UB}}}\frac{\sigma^{2}}{k}.
	\end{aligned}
\end{equation}
Meanwhile, the CDF of the upper bound on the maximum channel gain is given by
\begin{equation}\label{eq_channel_staCDFInfUB}
	\begin{aligned}
		F_{\infty}^{\mathrm{UB}}(t) =\Pr \left\{\max \limits_{1 \leq k \leq N_{\mathrm{UB}}} \tilde{X}_{k} \leq t \right\}= \left(1-e^{-\frac{t}{\sigma^{2}}}\right)^{N_{\mathrm{UB}}}.
	\end{aligned}
\end{equation}
Thus, for any given transmit power $p_{\mathrm{t}}$, noise power $\delta^{2}$, and receive SNR threshold $\gamma_{\mathrm{th}}$, a lower bound on the outage probability for the MA system with infinite number of receive paths is given by $F_{\infty}^{\mathrm{UB}}(\frac{\delta^{2}\gamma_{\mathrm{th}}}{p_{\mathrm{t}}})$. It is known that the sum of the first $N$ terms of the harmonic series can be approximated by $\sum \limits_{k=1}^{N}\frac{1}{k} \approx \log N + \varsigma$, with $\varsigma \approx 0.577$ denoting the Euler-Mascheroni constant. Thus, the lower and upper bounds in \eqref{eq_channel_stamaxInfLB} and \eqref{eq_channel_stamaxInfUB} indicate that the expected value for the maximum channel gain in the receive region increases with the region size approximately following a logarithmic function.


\subsection{Implementation Issues of MA Systems}
Note that the above analysis focuses on the common characteristics of MA-enabled communication systems by demonstrating the fundamental relations between the channel gain improvement and the number of channel paths as well as the region size for antenna movement. Next, we address some practical issues in implementation of MA systems, including operation frequency band, antenna movement, and antenna position optimization.

\subsubsection{Frequency Band}
It is noteworthy that the channel characteristics of MA-enabled communication systems can exhibit variations across different frequency bands, such as the distributions of AoD, AoA, and PRM. For instance, in low-frequency bands like sub-6 GHz, the environment's rich scattering tends to generate large number of multi-path components, thus leading to substantial channel spatial variation/diversity to be exploited by antenna movement. In contrast, high-frequency bands, such as mmWave frequency bands, may experience channel sparsity, which limits the number of multi-path components and potentially results in a reduced spatial diversity gain when utilizing MAs. Also note that the antenna dimension and the required size of regions for antenna movement to achieve a certain diversity gain are both proportional to the signal wavelength. Consequently, sub-6 GHz bands generally require larger regions for moving antennas as compared to mmWave frequency bands. 

\subsubsection{Antenna Movement}
The proposed MA-enabled communication system necessitates the integration of an antenna positioning module along with the conventional communication module at wireless communication Tx/Rx. One viable approach is by employing a step motor to facilitate the movement of the antenna along a slide track \cite{Zhuravlev2015experi,zhu2023MAMag}. This method enables precise control of the MA's position by the central processing unit (CPU), for accommodating the communication requirements. The motor-based MA system is particularly well-suited for large-scale implementations, as the driving power of the motor and the length of the slide track can be scaled up to match the antenna's weight and size. Alternatively, the implementation of MA can also leverage micro-electromechanical systems (MEMS), known for their miniaturization, low power consumption, and high linearity \cite{balanis2011modern}. MEMS-enabled MA offers high positioning accuracy, making it suitable for small-scale systems, such as mmWave and THz transceivers. For acquiring further insights into the implementation of MEMS-enabled MA, interested readers may refer to \cite[Chapter 17]{balanis2011modern}.

\subsubsection{Antenna Position Optimization}
In MA systems, how to efficiently find the MA positions for achieving the optimal communication performance is a crucial problem to solve. One practical approach is by discretizing the Tx/Rx region into multiple grids and applying an exhaustive search of all possible MA positions over these grids to measure the corresponding channels and thereby identify the best MA positions. In addition. an alternative strategy is to estimate the channel map between the Tx and Rx regions, which constitutes the channel response between any point in the Tx region and any point in the Rx region, based on the sparse channel measurements made by Tx/Rx MAs over a sufficient number of different location pairs \cite{ma2023channel}. Given the estimated channel map, various optimization techniques can be applied to determine the MA positions, such as successive convex approximation (SCA), gradient descent/ascent, and alternating optimization (AO) \cite{ma2022MAmimo,zhu2023MAmultiuser}. In practice, if the space of the Tx/Rx region is limited and the number of MAs is small, the exhaustive search-based approach is preferred for moving the antennas thanks to the limited number of candidate positions. However, for multiple MAs deployed in large-size Tx/Rx regions, this approach may incur prohibitively high time overhead as well as energy consumption for antenna movement. In such scenarios, the channel map-based optimization strategy turns to be a more suitable solution by developing efficient algorithms for channel map estimation and antenna position optimization.

\subsubsection{Time Cost}
The time cost of an MA system is mainly caused by moving the antenna from the current position to a target position, which is determined by the moving distance and speed of the MA. In practice, the antenna-moving speed may depend on the weight of the antenna, the output power of the mechanical driver, and the resistance of the MA slide. The time cost of antenna movement can be efficiently reduced by increasing the antenna-moving speed, e.g., improving the performance of the mechanical driver, or decreasing the antenna-moving distance, e.g., selecting a locally optimal position for the MA close to its current location. Nevertheless, the response time of MA systems, typically ranging from dozens of milliseconds to several seconds, is much larger than that of FPA systems with/without AS (usually in the order of microsecond or nanosecond). As such, the proposed MA system is more suitable to be used in low-mobility scenarios. It is also noted that antenna movement requires additional energy consumption, which can be kept low in practice by designing efficient antenna movement mechanism. For example, the MAs are only moved when the current antenna positions suffer from deep fading or the MAs' positions are optimized based on statistical CSI over a long time period.

\section{Simulation Results}
In this section, extensive simulations are carried out to evaluate the performance of the MA-enabled communication systems and verify our analytical results. 

\begin{figure*}[t]
	\centering
	\subfigure{\includegraphics[width=5.4 cm]{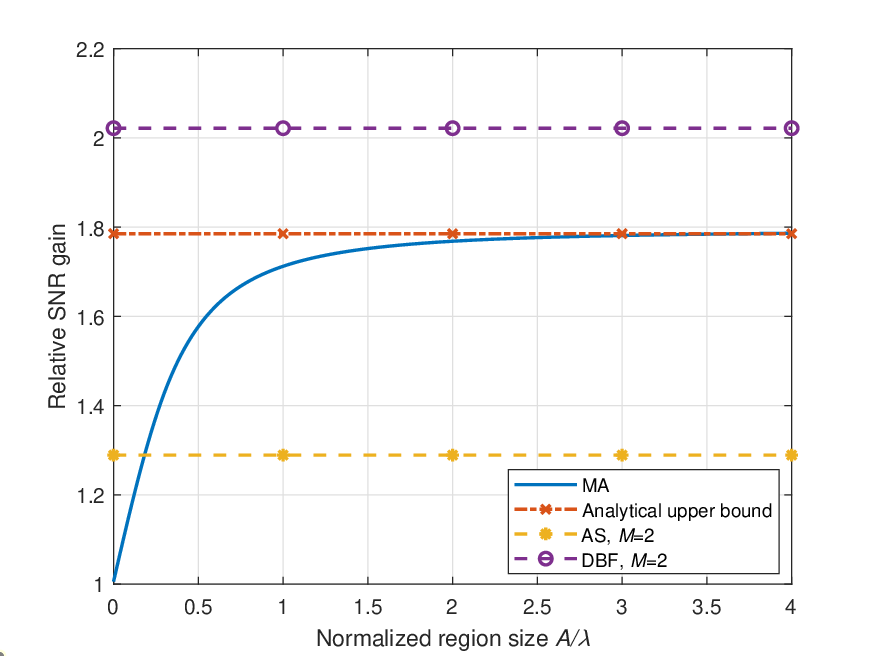}}
	\subfigure{\includegraphics[width=5.4 cm]{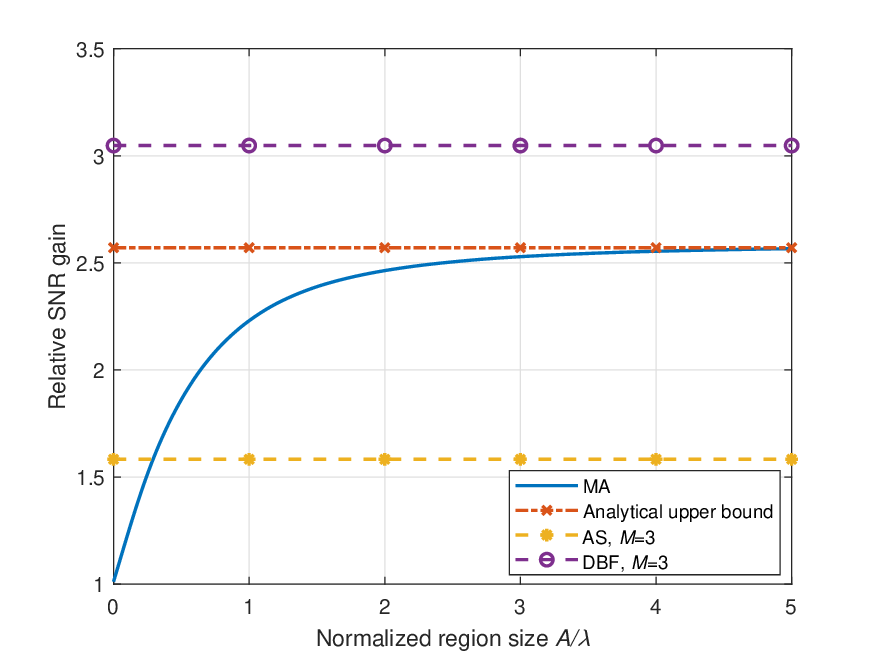}}
	\subfigure{\includegraphics[width=5.4 cm]{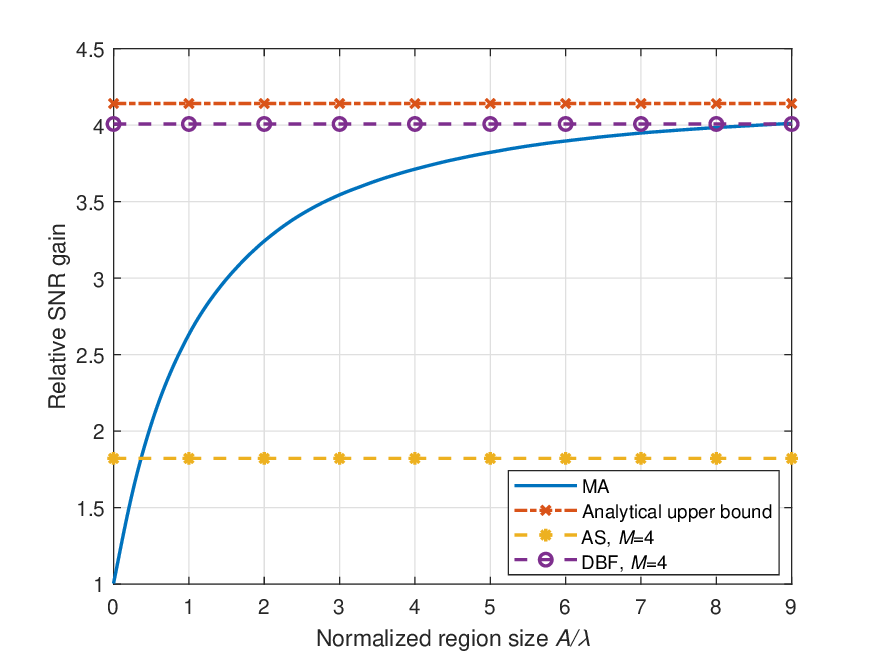}}
	\caption{Comparison of the relative SNR gains for MA, AS, and DBF systems versus the normalized size of the receive region with the number of receive paths $L_{\mathrm{r}}=2$, $L_{\mathrm{r}}=3$, and $L_{\mathrm{r}}=5$, respectively.}
	\label{Fig:SNRcom_SISO}
\end{figure*}

\begin{figure*}[t]
	\centering
	\subfigure{\includegraphics[width=5.4 cm]{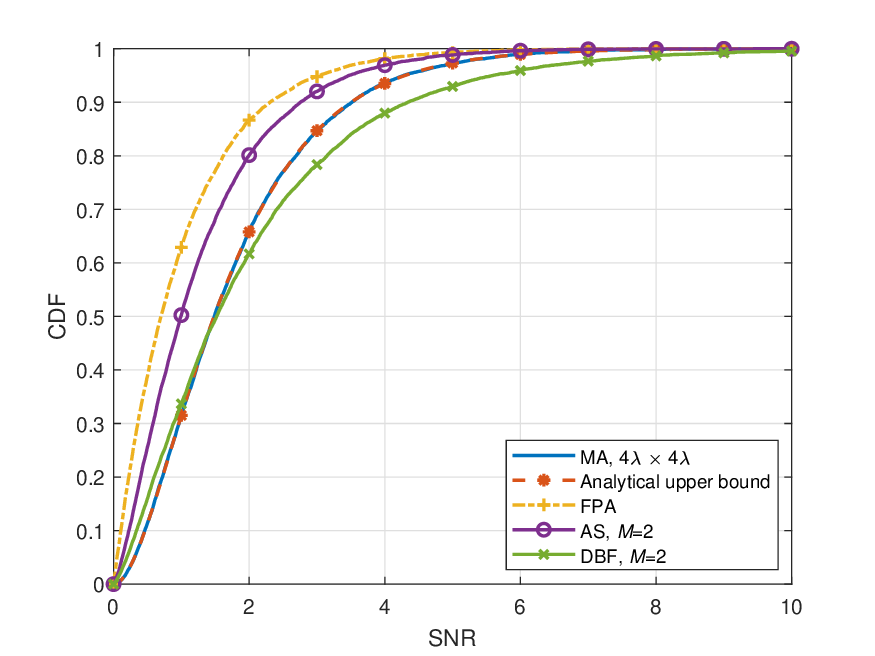}}
	\subfigure{\includegraphics[width=5.4 cm]{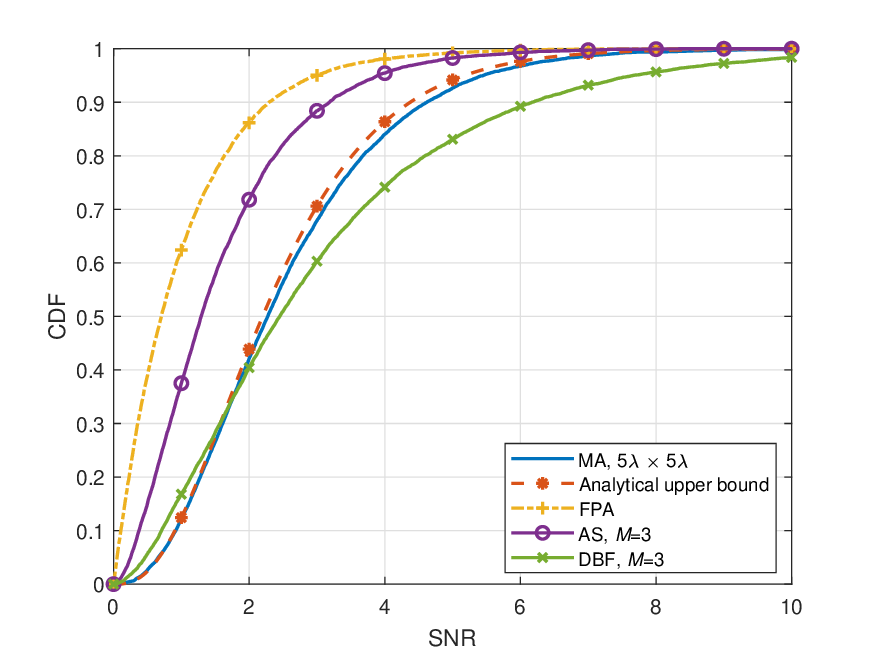}}
	\subfigure{\includegraphics[width=5.4 cm]{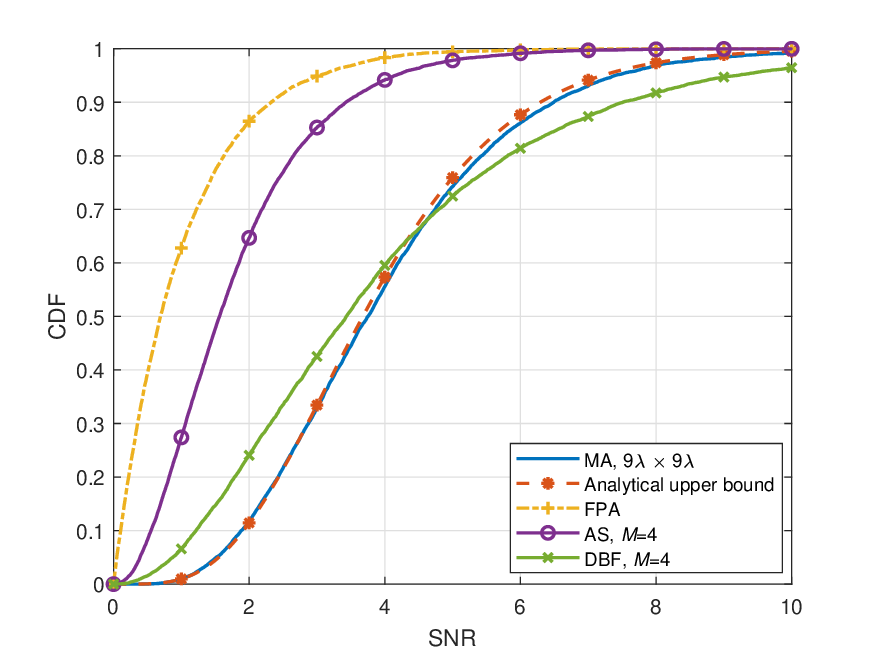}}
	\caption{Comparison of the SNR CDFs for MA, FPA, AS, and DBF systems with the number of receive paths $L_{\mathrm{r}}=2$, $L_{\mathrm{r}}=3$, and $L_{\mathrm{r}}=5$, respectively.}
	\label{Fig:SNRCDF_SISO}
\end{figure*}

\subsection{Simulation Setup and Benchmark Schemes}
In the simulation, the position of the transmit antenna is fixed, while the receive MA can be moved flexibly in the receive region, which is set as a square area with size $A \times A $, i.e., $\mathcal{C}_{\mathrm{r}}=[-A/2,A/2] \times [-A/2,A/2]$. The ratio of the average receive signal power to the noise power is set as $\frac{p_{\mathrm{t}}\sigma^{2}}{\delta^{2}}=1$ for convenience. The geometry channel model is employed as a special case of our proposed field-response based model, where the number of transmit and receive paths are the same, i.e., $L_{\mathrm{t}}=L_{\mathrm{r}}$, with a diagonal PRM $\mathbf{\Sigma} = \mathrm{diag}\{\sigma_{1}, \sigma_{2}, \cdots, \sigma_{L_{\mathrm{r}}}\}$. The path-response coefficients are assumed to be i.i.d. CSCG random variables, i.e., $\sigma_{\ell} \sim \mathcal{CN}(0,\sigma^{2}/L_{\mathrm{r}})$, $1\leq \ell \leq L_{\mathrm{r}}$. Besides, the physical AoDs and AoAs are assumed to be i.i.d. random variables with the PDF given by \eqref{eq_angle_Ray}. For each channel realization, we employ exhaustive search for the position of the receive MA which yields the highest channel gain.\footnote{The exhaustive search for the MA position is used to validate the analytical findings presented in this paper. In practice, efficient optimization techniques, such as SCA, gradient descent/ascent, and AO, can be employed for antenna position optimization based on the known/estimated channel state information (CSI) between the Tx and Rx regions \cite{zhu2023MAmultiuser,ma2022MAmimo}.} Specifically, the receive region is divided into multiple grids of equal size, $\lambda/100 \times \lambda/100$. The center of each grid serves as the reference point for measuring the channel gain of that grid. According to \eqref{eq_field_resRx}, it can be readily verified that the phase difference of each channel path's coefficient between the center and any point in the same grid does not exceed $\sqrt{2}\pi/100$. Thus, the discrete search method provides a near-optimal position of the MA for maximizing the channel gain. The results in this section are carried out based on $10^4$ Monte Carlo simulations. 

In addition to the proposed MA system, three benchmark schemes are defined as follows:
\begin{itemize}
	\item FPA: The antenna at the Rx has a fixed position located at the reference point $\mathbf{r}_{0}=(0,0)$.
	\item AS: The Rx is equipped with 1 RF chain and an array with $M$ antennas spaced by a half wavelength in the receive region. In particular, the antenna with the highest channel gain is selected out of the $M$ antennas for maximizing the receive SNR. 
	\item Digital beamforming (DBF): In the SIMO system, the Rx is equipped with $M$ RF chains and an array with $M$ antennas spaced by a half wavelength in the receive region. In particular, the maximum ratio combining (MRC) is employed at the Rx for maximizing the receive SNR.
\end{itemize}
For all the schemes, we define the the relative SNR gain as the ratio of the SNR for the considered scheme to the SNR for the scheme with a single transmit FPA at $\mathbf{t}_{0}$ and a single receive FPA at $\mathbf{r}_{0}$.

\subsection{Numerical Results}

Fig. \ref{Fig:SNRcom_SISO} shows the relative SNR gains for MA, AS, and DBF systems with varying normalized size of the receive region. The number of receive paths is set as 
$L_{\mathrm{r}}=2$, $L_{\mathrm{r}}=3$, and $L_{\mathrm{r}}=5$ for the three subplots, respectively. As can be observed, the relative SNR gain of the MA system increases with the region size because a higher maximum channel gain can be found in a larger receive region. The analytical upper bounds on the relative SNR gain are given by \eqref{eq_SNR_gain2}, \eqref{eq_SNR_gain3}, and \eqref{eq_SNR_gainL}. It can be seen that for the two-path and three-path cases, the upper bounds provided by \eqref{eq_SNR_gain2} and \eqref{eq_SNR_gain3} are tight due to the strong periodicity of the channel gain. The minimum size of the receive region for achieving the upper bound on the relative SNR gain is $4 \lambda$ and $5 \lambda$ for $L_{\mathrm{r}}=2$ and $L_{\mathrm{r}}=3$, respectively. This result is consistent with the derivation in \eqref{eq_channel_max_dis2} and \eqref{eq_channel_max_dis3}, where a lager size of the region is required to guarantee achieving the upper bound on the maximum channel gain as the number of receive paths increases. This conclusion can be further verified by the case of $L_{\mathrm{r}}=5$, where the relative SNR gain still increases if the region size is larger than $8 \lambda$. The reason is that as the number of paths increases, the periodicity becomes weaker in the receive region. According to the period analysis in \eqref{eq_channel_power_periodL}, the maximal common factor for the virtual AoA indices decreases with the number of paths, which entails a larger period for the channel gain in the receive region. Thus, to achieve the upper bound on the relative SNR gain, a larger size of the receive region is required for the case of more paths. Besides, the relative SNR gains of AS and DBF systems are also evaluated in Fig. \ref{Fig:SNRcom_SISO}. As can be observed, although the number of antennas for AS and DBF systems is $M$ times larger than that for the MA system, the MA system can achieve a performance comparable to that of the DBF system and better than that of the AS system. For the AS system, the relative SNR gain is small because the candidate antennas have fixed positions. In contrast, the MA system has more flexibility to adjust the position of the antenna for achieving a higher SNR. Note that the DBF system can acquire significant array gain because multiple receive antennas employ MRC for maximizing the receive SNR. However, the number of RF chains increases with the number of antennas, and thus the hardware cost for the DBF system is larger than that for the MA/AS system. As the number of channel paths increases, the proposed MA system can achieve higher diversity gain due to the more pronounced small-scale fading in the receive region. For example, if the number of channel paths is $L_{\mathrm{r}}=5$, the MA system can reap a relative SNR gain even higher than that of the DBF system with 4 FPA antennas. This is because by properly deploying the MA, the complex coefficients of multiple channel paths can be constructively combined at the Rx.

Fig. \ref{Fig:SNRCDF_SISO} compares the SNR CDFs for MA, FPA, AS, and DBF systems with the same setup as Fig. \ref{Fig:SNRcom_SISO}. As can be observed, the simulated CDF of the SNR for the two-path MA system perfectly matches the analytical result derived in \eqref{eq_channel_staCDF2}. While for the $L_{\mathrm{r}}=3$ and $L_{\mathrm{r}}=5$ cases, the simulated CDF has a deviation compared to the analytical ones because an approximation was employed in \eqref{eq_channel_staCDF3} and \eqref{eq_channel_staCDFL}. Nevertheless, the error between the analytical and simulated CDFs is very small, which indicates that \eqref{eq_channel_staCDF3} and \eqref{eq_channel_staCDFL} provide a good approximation for the CDF of the upper bound on the maximum channel gain and can be utilized for analyzing the outage probability for the MA system. Besides, we can find that for any fixed SNR threshold, the value of the CDF for the MA system is always smaller than that for FPA and AS systems, which means that compared to the two benchmark schemes, the MA system can always achieve a lower outage probability for any SNR threshold. Interestingly, at the low SNR region, the SNR CDF value for the MA system is smaller than that for the DBF system. It indicates that the MA system can outperform the DBF system in terms of the outage probability at the low SNR region. In particular, for a large number of receive paths, the performance gap of the outage probability increases, which shows more superiority of the MA system. This is attributed to the fact that the MA can consistently identify a point that yields locally maximal channel gain within a given receive region. In contrast, the conventional FPA may be located at a position with poor channel conditions. In the most unfavorable scenario, even if multiple antennas are utilized, they may all experience deep fading channels at fixed locations (e.g., the local area with low channel gains depicted in Fig. \ref{fig:PowerChannelL4}).

\begin{figure}
	\centering
	\includegraphics[width=8 cm]{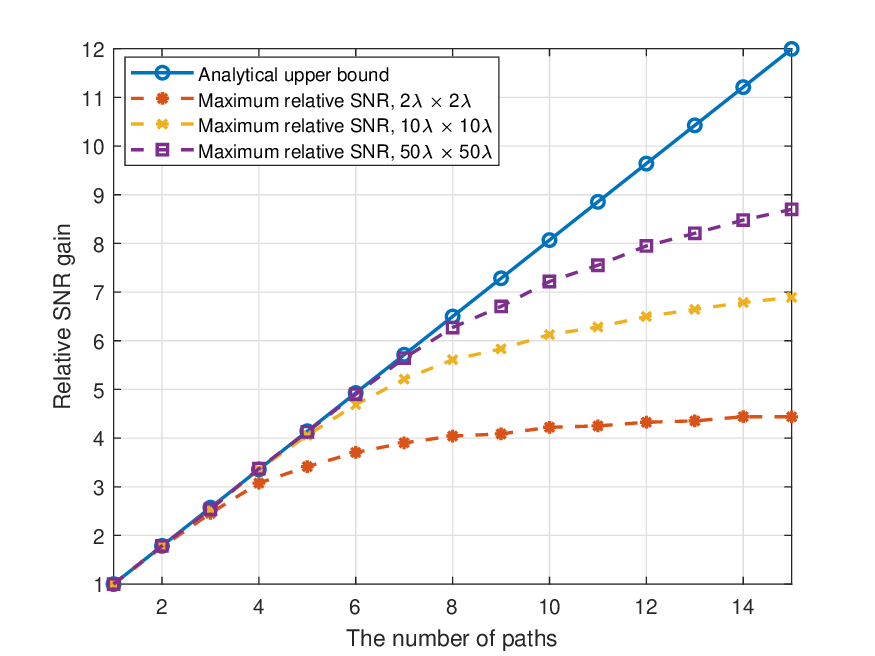}
	\caption{The expected values of the relative SNR gains for MA systems versus the number of receive paths for different sizes of the receive region.}
	\label{fig:NumPath}
\end{figure}

\begin{figure}
	\centering
	\includegraphics[width=8 cm]{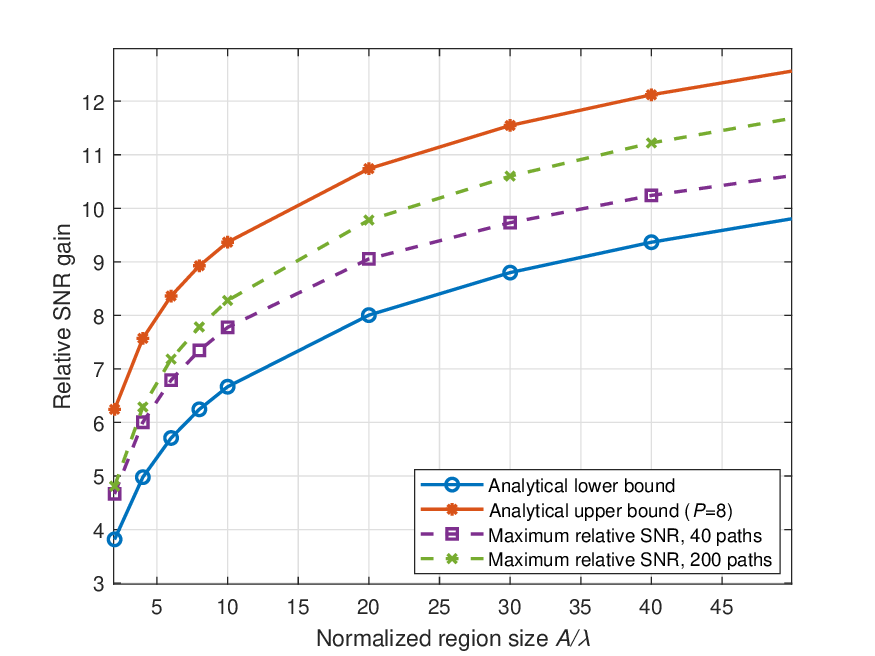}
	\caption{The expected values of the relative SNR gains versus the normalized size of the receive region for different numbers of receive paths.}
	\label{fig:RegionSize}
\end{figure}

In Fig. \ref{fig:NumPath}, we compare the relative SNR gain and the upper bound on the maximum SNR gain for the MA system with varying number of receive paths. As can be observed, both the relative SNR gain and the upper bound on the maximum SNR gain increase with the number of paths. This is because the small-scale fading becomes stronger as the number of paths increases, and thus the channel gain has more substantial fluctuation in the receive region, which entails an increase on the maximum channel gain. Besides, as the region size increases from $2 \lambda \times 2 \lambda$ to $50 \lambda \times 50 \lambda$, the maximum SNR gain closely approaches the analytical upper bound provided by \eqref{eq_SNR_gainL}. In particular, if the number of paths is smaller than 3, the performance gap between the maximum SNR gain for $2 \lambda \times 2 \lambda$ region and the upper bound is no more than 0.12. If the number of paths is smaller than 5, the performance gap between the maximum SNR gain for $10 \lambda \times 10 \lambda$ region and the upper bound is no more than 0.10. If the number of paths is smaller than 7, the performance gap between the maximum SNR gain and the upper bound is no more than 0.08. The above results indicate that the upper bound in \eqref{eq_SNR_gainL} is approximately tight for a small number of paths. For larger number of paths, a larger size of the receive region is required to achieve the SNR performance near to the upper bound. It is worth noting that the upper bound is effective for infinite size of the receive region. It is expected that the maximum SNR gain can further increase if the size of the receive region is larger than $50 \lambda \times 50 \lambda$ and approach the upper bound for infinitely large size of the receive region more closely.

Fig. \ref{fig:RegionSize} shows the expected value of the relative SNR gain versus the normalized size of the receive region as well as the comparison with the lower and upper bounds in \eqref{eq_channel_stamaxInfLB} and \eqref{eq_channel_stamaxInfUB}. The number of paths is set as 40 or 200, and the discretized parameter for the upper bound is set as $P=8$ in \eqref{eq_channel_stamaxInfUB}. As can be observed, the relative SNR gains increase with the size of the region, and the performance gap to the upper bound decreases if the number of paths increases. As we have analyzed in Section III-B, the lower and upper bounds in \eqref{eq_channel_stamaxInfLB} and \eqref{eq_channel_stamaxInfUB} both approximately follow a logarithmic function with the region size. The results in Fig. \ref{fig:RegionSize} validate this analysis and show that the relative SNR gains for sufficiently large number of paths have a logarithmic growth with respect to the size the receive region.

\begin{figure*}[t]
	\centering
	\subfigure{\includegraphics[width=5.4 cm]{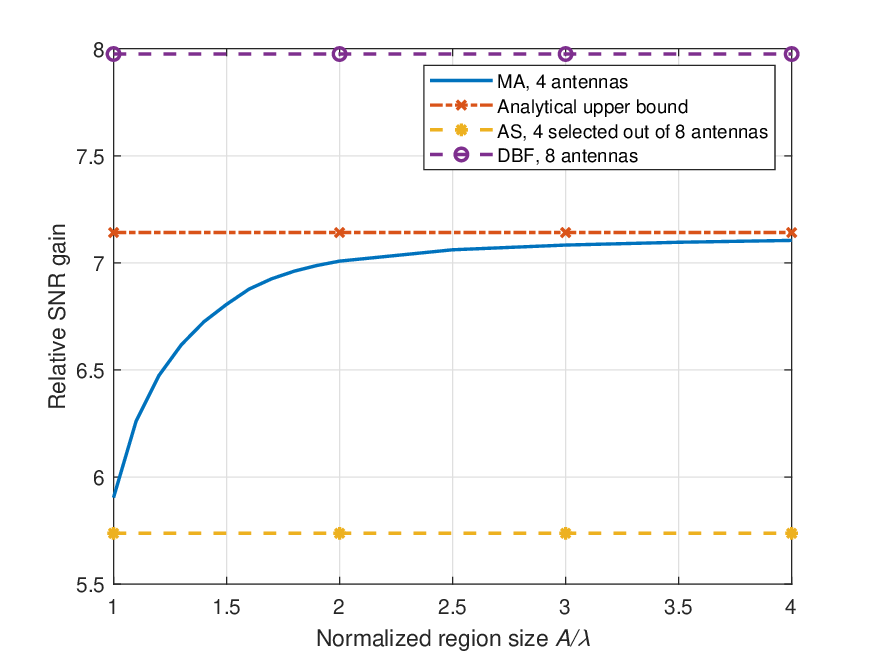}}
	\subfigure{\includegraphics[width=5.4 cm]{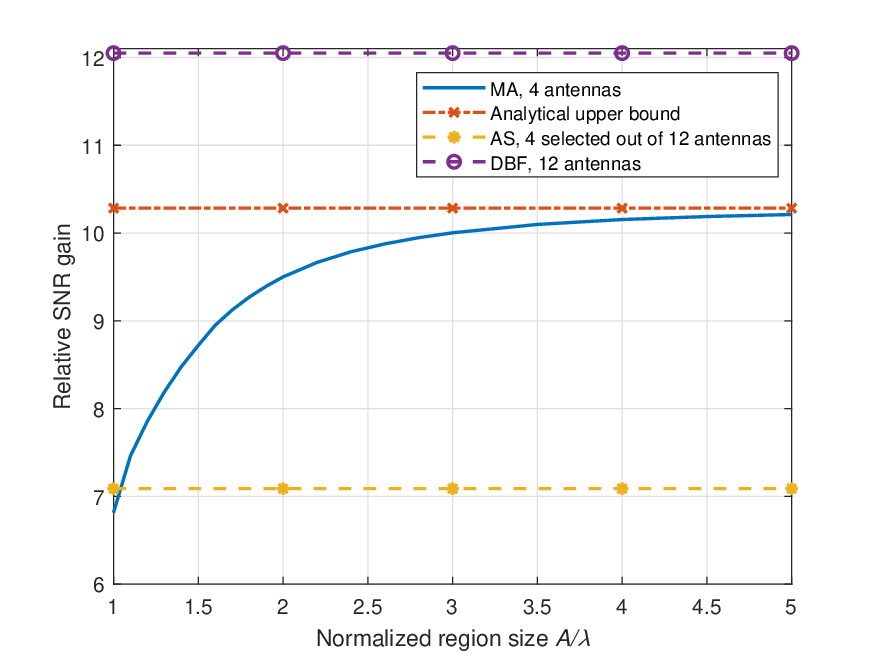}}
	\subfigure{\includegraphics[width=5.4 cm]{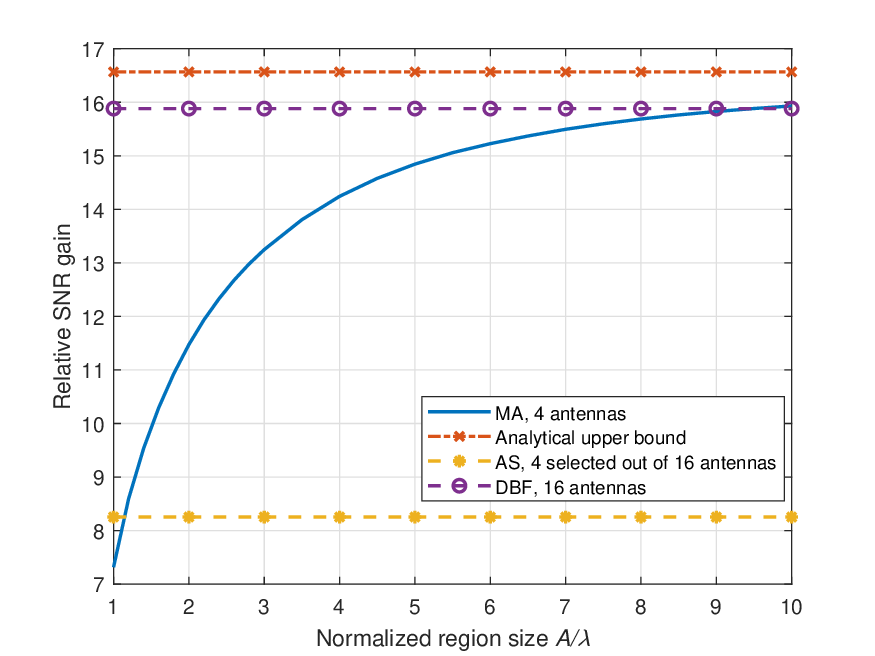}}
	\caption{Comparison of the relative SNR gains for multi-MA, AS, and DBF systems versus the normalized size of the receive region with the number of receive paths $L_{\mathrm{r}}=2$, $L_{\mathrm{r}}=3$, and $L_{\mathrm{r}}=5$, respectively.}
	\label{Fig:SNRcom_SIMO}
\end{figure*}

Next, we evaluate the relative SNR gains of multiple MAs with varying normalized size of the receive region and compare them with the AS and DBF systems in Fig. \ref{Fig:SNRcom_SIMO}. The number of MAs is set as 4, where the positions having the first four highest channel gains are sequentially searched in the receive region, subject to the constraint that the distance between any two MAs is no less than half wavelength. The DBF and AS systems both employ $4M$ antennas, where 4 antennas with the highest channel gains are selected out of the candidate antennas for the AS system. For all schemes, MRC is utilized at the Rx for maximizing the receive SNR. The analytical upper bounds on the relative SNR gain for MAs are given by \eqref{eq_SNR_gain2}, \eqref{eq_SNR_gain3}, and \eqref{eq_SNR_gainL}, multiplied by the number of antennas, i.e., 4. It can be observed again that the relative SNR gain of MAs increases with the region size and approaches the analytical upper bound. As we have shown in Section III-A, the channel gain exhibits a periodic character for the cases of $L_{\mathrm{r}}=2$ and $L_{\mathrm{r}}=3$. Thus, the multiple MAs can be deployed at different positions which achieve the maximum channel gain as long as the size of the receive region is properly enlarged. Moreover, for the case of $L_{\mathrm{r}}=5$, the multiple MAs can still be deployed at the positions with local maximum channel gains, although the channel-gain pattern does not have an explicit period in the receive region. Besides, the MA system can always outperform the AS for all cases and achieve a comparable relative SNR gain with the DBF system utilizing more number of FPAs. In short, the results in Fig. \ref{Fig:SNRcom_SIMO} reveal that the spatial diversity gain of single MA is extendable to multi-MA systems.

\subsection{Performance Evaluation Based on 3GPP Channel Model}
To verify the performance gain of MA in practical systems, we conduct simulations based on the widely used 3GPP wideband channel model \cite{3gpp2019studyo}. In particular, we consider the typical scenario of indoor factory with dense clutter and high BS (InF-DH)\footnote{The future smart factory will be endowed with automation and intelligence for reducing manual operation cost, where a large number of machine-type terminals are usually deployed at fixed locations or with low mobility, and their surrounding propagation environment typically varies slowly. In such scenarios, MAs can be installed on these terminals, such as machines, low-mobility vehicles and robots, for enhancing their communication performance.} of size $20 \times 20 \times 10~\text{m}^{3}$, where the antenna heights for the BS and user terminal (UT) are set as 10 m and 1.5 m, respectively. The 2D horizontal distance between the BS and UT is fixed as 20 m. In the considered scenario, 3GPP recommends the use of 25 clusters for both LoS and NLoS channels, each consisting of 20 rays per cluster. For the MA system, the BS employs an FPA transmitting orthogonal frequency division multiplexing (OFDM) symbols to the UT equipped with an MA, where the region for antenna movement is extended to a 3D space of size $A \times A \times A$.  For the AS system, the antennas are overspread in the entire 2D surface of size $A \times A$, resulting in the number of antennas scaling with the region size. More precisely, with the half-wavelength spacing between adjacent antennas, the total number of antennas in a $t\lambda \times t\lambda$ region is given by $(2t+1)^2$, $t=0,1,2,\cdots,10$. The setups for the FPA and DBF systems are similar to that specified in Section IV-A. For all systems, we select the carrier frequency of 5.2 GHz and the bandwidth of $B=20$ MHz. The noise power density is set as $N_{0}=-174$ dBm/Hz and the total transmit power of the BS is $P_{\mathrm{t}} = 15$ dBm, which is equally allocated to all subcarriers. The numbers of subcarriers and cyclic prefix (CP) are set as $N=64$ and $N_{\mathrm{CP}}=8$, respectively. Denoting the channel frequency response (CFR) for the $n$-th subcarrier at MA position $\mathbf{r}$ as $h_{n}(\mathbf{r})$, $1 \leq n \leq N$, we obtain the achievable rate of the OFDM system as 

\begin{equation}\label{eq_rate_OFDM}
	\begin{aligned}
		R(\mathbf{r}) = \frac{1}{N+N_{\mathrm{CP}}} \sum \limits_{n=1}^{N} \log_{2} \left( 1+\frac{|h_{n}(\mathbf{r})|^{2}P_{\mathrm{t}}}{BN_{0}} \right),
	\end{aligned}
\end{equation}
where $P_{\mathrm{t}}$ and $N_{0}$ are converted from dB to absolute values.

\begin{figure}
	\centering
	\includegraphics[width=8 cm]{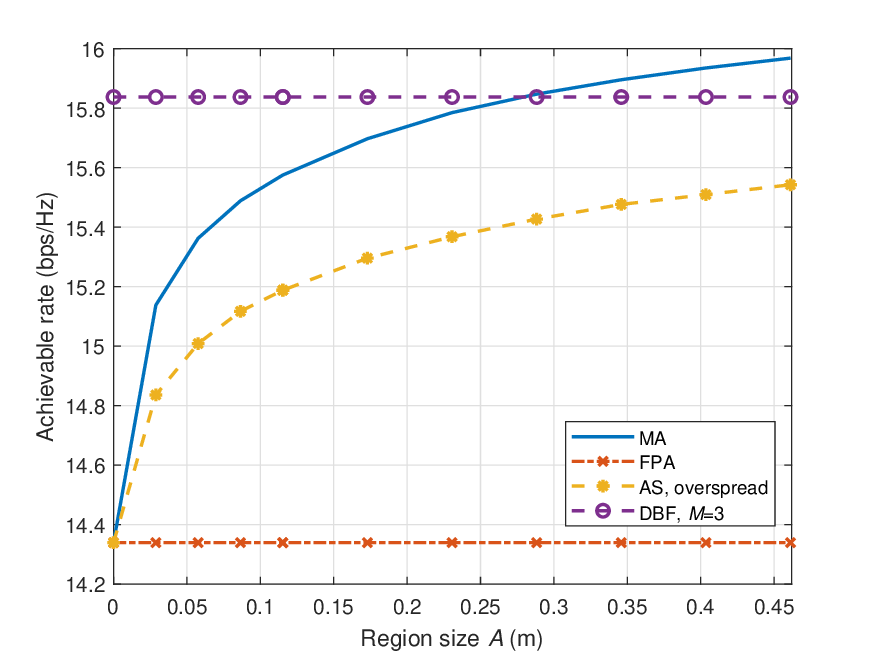}
	\caption{Achievable rates of MA, FPA, AS, and DBF systems under the 3GPP LoS wideband channel model.}
	\label{fig:3GPP_LoS}
\end{figure}

\begin{figure}
	\centering
	\includegraphics[width=8 cm]{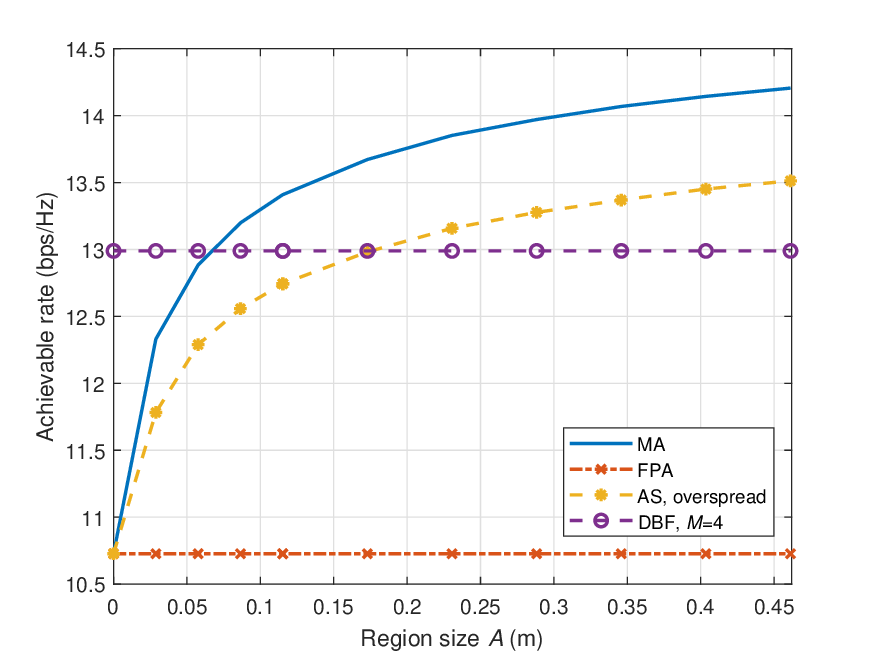}
	\caption{Achievable rates of MA, FPA, AS, and DBF systems under the 3GPP NLoS wideband channel model.}
	\label{fig:3GPP_NLoS}
\end{figure}

Figs. \ref{fig:3GPP_LoS} and \ref{fig:3GPP_NLoS} depict the achievable rates of OFDM systems aided by MA, FPA, AS, and DBF under LoS and NLoS channel models, respectively. It is observed from the figures that the effective (maximum) achievable rate of the MA system exhibits an increasing trend with the region size for both LoS and NLoS channels. This is because the MA position optimization can increase the average power of the CFR as well as improve the power distribution among multiple subcarriers. Note that a large region size requires an extended space for installing the MA on the UT. Thus, a trade-off exists between increasing the performance gain of MA systems and decreasing the region size for moving antenna. In practice, an appropriate size of the antenna-moving region should be selected according to the carrier frequency, hardware cost constraints, and communication requirements. Besides, the results in Figs. \ref{fig:3GPP_LoS} and \ref{fig:3GPP_NLoS} show that MA system outperforms the conventional FPA system and can even achieve comparable performance to the DBF system. Notably, for a region size of $A=0.1$ m, the achievable rate improvement over the FPA system reaches up to 1.1 bps/Hz and 2.6 bps/Hz for LoS and NLoS channels, respectively. The better improvement with NLoS channels can be attributed to NLoS channels having a more pronounced small-scale fading due to larger number of channel paths and more balanced distribution of channel path powers. Although the AS scheme can also enhance the achievable rate as the region size increases, it incurs significantly higher antenna cost because the entire receive region is covered with antennas. For instance, a region size of $A=0.46$ m would require 289 antennas for AS. Moreover, due to the antenna layout constraint, the antennas in the AS scheme can only be placed on a 2D surface. In contrast, the MA system offers greater flexibility by allowing a single antenna to move in a 3D space, thereby achieving higher spatial diversity gains.

\section{Conclusion}
In this paper, we proposed a new MA architecture for improving the performance of wireless communication systems. With the capability of flexible movement, the MAs can be deployed at positions with more favorable channel conditions in the spatial region to achieve higher spatial diversity gains. Since the channel varies with the positions of the MAs, a field-response based channel model was developed under the far-field condition to characterize the general multi-path channel with given transmit and receive regions. We have shown the conditions under which the proposed field-response based channel model for MAs becomes the well-known LoS channel, geometry channel, Rayleigh and Rician fading channel models with FPAs. Based on the proposed channel model, we analyzed the maximum channel gain achieved by a single receive MA over its FPA counterpart under both deterministic and stochastic channels. In the deterministic channel case, the periodic behavior of the multi-path channel gain was revealed in a given spatial field, which unveiled the maximum channel gain of the MA with respect to the number of channel paths and size of the receive region. In the case of stochastic channels, we derived the expected value of an upper bound on the maximum channel gain of the MA in an infinitely large receive region for different numbers of channel paths. Moreover, our results revealed that higher performance gains by the MA over the FPA could be acquired when the number of channel paths increases due to more pronounced small-scale fading effects in the spatial domain. Besides, for uniform scattering scenarios with infinite number of channel paths, we showed that the upper and lower bounds on the expected maximum channel gain of the MA obey the logarithmic increase with respect to the size of the receive region. The approximate CDF of the maximum channel gain was also obtained in closed form, which is useful to evaluate the outage probability of the MA system. Simulation results validated our analytical results and demonstrated that the MA system could reap considerable performance gains over the conventional FPA systems with/without AS, and even achieve comparable performance to the SIMO beamforming system. With the general channel model and theoretical performance limits of the MA system provided in this paper, the design of practical methods for achieving the performance limits, such as channel estimation and antenna position searching algorithms, as well as the extension of the results to multi-antenna/multi-user systems with spatial multiplexing will be interesting topics for future research. Besides, the performance analysis and practical design for wideband MA systems are important problems requiring further investigation.

\bibliographystyle{IEEEtran} 
\bibliography{IEEEabrv,ref_zhu}

\end{document}